\documentclass[a4paper,11pt]{article}
\usepackage{amssymb}
\usepackage{graphicx}
\usepackage{amsmath}

\DeclareFontFamily{U}{rsf}{}
\DeclareFontShape{U}{rsf}{m}{n}{
  <5> <6> rsfs5 <7> <8> <9> rsfs7 <10-> rsfs10}{}
\DeclareMathAlphabet\Scr{U}{rsf}{m}{n} \makeatletter
\@addtoreset{equation}{section} \makeatother

\def\be{\begin{equation}}
\def\ee{\end{equation}}
\def\ba{\begin{array}}
\def\ea{\end{array}}

\newcommand{\bea}{\begin{eqnarray}}
\newcommand{\eea}{\end{eqnarray}}
\newcommand{\Spin}{\mathrm{Spin}}

\newcommand{\al}{\alpha}

\usepackage{color}

\newcommand{\R}{\mathbb R}
\newcommand{\C}{\mathbb C}
\newcommand{\Z}{\mathbb Z}
\newcommand{\bet}{\beta}
\newcommand{\rGL}{\mathrm{GL}}
\newcommand{\Hom}{\mathrm{Hom}}
\newcommand{\salg}{\mathrm{(salg)}}
\newcommand{\lra}{\longrightarrow}

\begin{document}

\begin{titlepage}

\begin{flushright}
DFPD/2016/TH/1
\end{flushright}

\vskip 2.0 cm
\begin{center}  {\huge{\bf      Klein and Conformal Superspaces, \\
     \vspace{5pt}Split Algebras and Spinor Orbits
    }}

\vskip 1.5 cm

{\Large{\bf Rita Fioresi$^{1}$}, {\bf Emanuele Latini$^1$}, {\bf Alessio Marrani$^{2,3}$}}

\vskip 1.0 cm

$^1${\sl Dipartimento di Matematica, Universit\`{a} di
Bologna\\Piazza di Porta S. Donato 5, I-40126 Bologna, Italy}\\
\texttt{rita.fioresi@UniBo.it}, \texttt{emanuele.latini@UniBo.it}

\vskip 0.5 cm

$^2${\sl Museo Storico della Fisica e Centro Studi e
Ricerche ``Enrico Fermi" \\ Via Panisperna 89A, I-00184, Roma, Italy}\\

\vskip 0.5 cm

$^3${\sl Dipartimento di Fisica e Astronomia ``Galileo
Galilei'', Universit\`a di Padova
\\ and INFN, Sez. di Padova \\
Via Marzolo 8, I-35131 Padova, Italy}
\\ \texttt{Alessio.Marrani@pd.infn.it}

 \end{center}

 \vskip 3.5 cm

\begin{abstract}
We discuss $\mathcal{N}=1$ Klein and Klein-Conformal superspaces in $D=(2,2)$
space-time dimensions, realizing them in terms of their functor of points
over the split composition algebra $\mathbb{C}_{s}$. We exploit the
observation that certain split forms of orthogonal groups can be realized in
terms of matrix groups over split composition algebras; this leads to a
natural interpretation of the the sections of the spinor bundle in the critical split
dimensions $D=4$, $6$ and $10$ as $\mathbb{C}_{s}^{2}$, $\mathbb{H}_{s}^{2}$
and $\mathbb{O}_{s}^{2}$, respectively. Within this approach, we also analyze the non-trivial spinor orbit stratification that is relevant in our construction since it affects the Klein-Conformal superspace
structure. \end{abstract}
\vspace{24pt} \end{titlepage}


\newpage

\tableofcontents

\newpage

\section{\label{Intro}Introduction}

\textsl{Supersymmetry} (\textsl{Susy}) is a deep and elegant symmetry
relating half-integer spin fields (\textsl{fermions}, constituents of
matter) to integer-spin fields (\textsl{bosons}, giving rise to
interactions). Such a symmetry was originally formulated, as a \textsl{global%
} symmetry of fields, back in the early 70's in former Soviet Union by
physicists Gol'fand and Likhtman \cite{GL}, Volkov and Akulov \cite{VA}, and
independently in Europe by Wess and Zumino \cite{WZ}.

A major advance in the formulation of supersymmetric theories in space-time,
which then allowed for the construction of manifestly invariant
interactions, was due to Salam and Strathdee, who were the first to
introduce the concept of \textsl{superfield} \cite{SS-1,SS-2}. In fact,
depending on the number $s$ and $t$ of \textsl{spacelike} resp. \textsl{%
timelike} dimensions, space-time Susy recasts bosonic and fermionic fields
into multiplet structures, each providing a certain representation of such
an underlying symmetry. Within the simplest formulation of Susy, in which a
unique fermionic generator exists besides the bosonic ones, fields defined
in a space $\mathbf{M}^{s,t}\cong \mathbb{R}^{s,t}$ (which in the case $s=3$
and $t=1$ yields the usual Minkowski space-time) are assembled into a unique
object, named \textsl{superfield}, defined into the so-called $\mathcal{N}=1$%
, $\left( s+t\right) $-dimensional superspace $\mathbf{M}^{s,t|1}$, which is
characterized by the presence of an anti-commuting Grassmannian coordinate
besides the usual commuting bosonic coordinates of $\mathbf{M}^{s,t}$.

Such developments eventually led to major advances in Quantum Field Theory,
constituting the foundational pillars on which consistent candidates for a
unified theory encompassing Quantum Gravity and the Standard Model of
particle interactions were constructed. In combination with local gauge
invariance, global Susy allowed for the formulation of Supersymmetric
Yang-Mills Theories (SYM's) \cite{FZ}. In such a framework, Susy gives rise
to remarkable cancellations between bosons and fermions in their quantum
corrections, thus allowing for a study of SYM's beyond perturbation theory.
This generally provides a framework for a possible solution of the \textsl{%
hierarchy problem}, for the search of natural candidates for \textsl{dark
matter}, as well as for addressing the conceptual issue of the \textsl{dark
energy}.

In presence of general diffeomorphisms covariance, Susy becomes a \textsl{%
local} symmetry. In 1976, Ferrara, Freedman, Van Niewenhuizen \cite{FFVN}
and Deser and Zumino \cite{DZ} succeeded in formulating Susy as a local
symmetry and coupling it to General Relativity. This resulted into the first
formulation of \textsl{supergravity}, providing a low-energy effective
description of more fundamental theories such as \textsl{superstrings} and ${%
M}$\textsl{-theory}, and playing a crucial role in \textsl{supersymmetry
breaking}, an essential ingredient of all realistic elaborations beyond the
Standard Model.

Also in its world-sheet formulation, Susy is one of the main tools for the
construction of the most promising frameworks - the aforementioned
superstring theory and $M$-theory - in which Quantum Theory and General
Relativity may be reconciled and consistently formulated (\textit{cfr. e.g.}
\cite{GSW-book,Polchinski-book}). Quite recently, local Susy also proved to
be a surprisingly successful tool in the investigation of the properties and
dynamics of \textsl{black holes}, the endpoints of gravitational collapse,
in which an horizon surface acts as a cosmic censor for the possible
formation of a space-time singularity.\bigskip

Susy had a major impact in Mathematics, as well (\textit{cfr.} \cite{vsv} for an excellent introduction). It gave rise
also to a vast, deep and flourishing arena of mathematical investigation,
inspiring generations of mathematicians to change their approach to
geometry, both from the differential and algebraic point of view. In such
frameworks, the symmetries of superspaces are naturally described by \textsl{%
superalgebras} and \textsl{supergroups}, the super-generalizations of the
usual concept of algebras and groups.

Nowadays, superseding the more traditional sheaf theoretic approach,
supergroups and \textsl{superspaces} are investigated by exploiting the
elegant machinery of the \textsl{functor of points}, originally introduced
by Grothendieck in algebraic geometry (see \textit{e.g.} \cite{bcf1,bcf2}).
Remarkably, such a deeply abstract point of view, formalized and developed
by Shvarts \cite{sh} and Voronov \cite{vo}, shares surprising similarities
with the physicists' approach in the aforementioned early times of Susy, in
which points in the superspace were understood by exploiting Grassmann
algebras, which are nothing but superalgebras over a superspace consisting
of a point \cite{be}. The subsequent work of Manin \cite{ma1,ma2}\ applied
the powerful abstract machinery of the functor of points to the theory of
superspaces and \textsl{superschemes}; ultimately, this led to the
development of the theory of \textsl{superflags} and \textsl{%
super-Grassmannians}.

However, sharing the same approach as in \cite{FL-1} and essentially relying
on \cite{ccf} and \cite{flv}, we would like to point out that in the present
investigation we will strive to leave abstract subtleties pertaining to the
formal machinery of functor of points on the background, though employing
its descriptive power while dealing with $T$-points of a supergroup or with
a superspace.\bigskip

An intriguing aspect of Susy is its deep relation to the four normed \textsl{%
division} algebras \cite{Hurwitz} $\mathbb{A}=\mathbb{R}$ (\textsl{real
numbers}), $\mathbb{C}$ (\textsl{complex numbers}), $\mathbb{H}$ (\textsl{%
quaternions}, or Hamilton numbers), $\mathbb{O}$ (\textsl{octonions}, or
Cayley numbers), especially involving \textsl{super-twistors} \cite%
{Baez-Huerta-1,cederwall1,cederwall2,cederwall3}. In fact, non-Abelian YM
theories are supersymmetric (thus giving rise to SYM's) only if the
space-time dimension is $D=3$, $4$, $6$ or $10$ (and the same is true for
the Green-Schwarz superstring), named \textsl{critical} dimension. In this
context, the consistent formulation of Susy relies on the vanishing of a
certain trilinear expression relying on the existence of $\mathbb{A}$, whose
real dimension is respectively given by $D-2$ \cite%
{evans,Baez-Huerta-1,Huerta-2,Huerta-3,Huerta}.

Motivated by attempts at explaining the remarkable fact that (super)gravity
scattering amplitudes can be obtained from those of (S)YM theories (\textit{%
cfr. e.g.} \cite{Bern}), in \cite{ICL-1309} Duff and collaborators exploited
normed division algebras $\mathbb{A}$'s in order to obtain the massless
spectrum and the multiplet structure of supergravity theories in various
dimensions by tensoring SYM multiplets (also \textit{cfr.} subsequent
developments in \cite{ICL-2,ICL-3}). The core of their main argument relies
on the observation that the entries of second row of the order-$2$ \textsl{%
split} magic square $\mathcal{L}_{2}\left( \mathbb{A}_{s},\mathbb{B}\right) $
\cite{MS,MS2,BS-2}\vspace{1mm}%
\begin{equation}
\begin{tabular}{ccccc}
\hline
&  &  &  &  \\[-4mm]
& $\mathbb{R}$ & $\mathbb{C}$ & $\mathbb{H}$ & $\mathbb{O}$ \\[0.1mm] \hline
&  &  &  &  \\[-3mm]
$\mathbb{C}_{s}$ & $\mathfrak{so}(2,1)$ & $\mathfrak{so}(3,1)$ & $\mathfrak{%
so}(5,1)$ & $\mathfrak{so}(9,1)$%
\end{tabular}%
\end{equation}%
can be naturally represented as $\mathfrak{sl}(2,\mathbb{A})$, then yielding
the isomorphisms of Lie algebras (\textit{cfr.} \cite{Baez}, as well as \cite%
{vsv-2,ICL-1309} and Refs. therein)%
\begin{equation}
\mathfrak{sl}(2,\mathbb{A})\cong \mathfrak{so}(q+1,1),  \label{iso-Lie-1}
\end{equation}%
where
\begin{equation}
q:=\text{dim}_{\mathbb{R}}\mathbb{A}=1,2,4,8\text{~for~}\mathbb{A}=\mathbb{R}%
,\mathbb{C},\mathbb{H},\mathbb{O}\text{,~respectively},  \label{q-def}
\end{equation}%
and $\mathfrak{so}(q+1,1)$ is the Lie algebra of the Lorentz group in $D=q+2$
dimensions. Analogously, the third line of $\mathcal{L}_{2}\left( \mathbb{A}%
_{s},\mathbb{B}\right) $, \textit{i.e.}\vspace{1mm}%
\begin{equation}
\begin{tabular}{ccccc}
\hline
&  &  &  &  \\[-4mm]
& $\mathbb{R}$ & $\mathbb{C}$ & $\mathbb{H}$ & $\mathbb{O}$ \\[0.1mm] \hline
&  &  &  &  \\[-3mm]
$\mathbb{H}_{s}$ & $\mathfrak{so}(3,2)$ & $\mathfrak{so}(4,2)$ & $\mathfrak{%
so}(6,2)$ & $\mathfrak{so}(10,2)$%
\end{tabular}%
\end{equation}%
can be reinterpreted by noting the following Lie algebraic isomorphism \cite%
{BS-2}%
\begin{equation}
\widetilde{\mathfrak{sp}}(4,\mathbb{A})\cong \mathfrak{so}(q+2,2),
\label{iso-Lie-2}
\end{equation}%
with $\mathfrak{so}(q+2,2)$ standing for the conformal Lie algebra in $D=q+2$%
, and $\widetilde{\mathfrak{sp}}(4,\mathbb{A})$ denoting the \textsl{%
Barton-Sudbery symplectic algebra}, in which the matrix transposition is
replaced by the Hermitian conjugation, differently from the usual definition
of symplectic algebras \cite{BS-1,BS-2}. The Lie algebraic isomorphisms (\ref%
{iso-Lie-1})-(\ref{iso-Lie-2}) have been recently extended to the Lie group
level (considering the spin covering of the Lorentz and conformal groups,
namely $\mathrm{Spin}(q+1,1)$ resp. $\mathrm{Spin}(q+2,2)$), by explicit
constructions worked out in a series of paper \cite%
{MS-2-Groups,Dray-2,Dray-3,Dray-4} by Dray, Manogue and collaborators. In
particular, in \cite{MS-2-Groups} a Lie group version of the aforementioned
order-$2$ split magic square $\mathcal{L}_{2}\left( \mathbb{A}_{s},\mathbb{B}%
\right) $ was constructed and studied.\bigskip

\textsl{Conformal symmetry} also plays a crucial role in Physics and in
Mathematics. While it is usually associated to massless particles, it also
characterizes, possibly as an approximated symmetry, a number of physical
systems in certain regimes of their dynamics.

Conformal symmetry also provides the foundation of an important branch of
geometry, named \textsl{conformal geometry}, in which equivalence classes of
metrics are exploited for a manifest, locally Weyl-invariant formulation of
the equations governing the evolution of physical systems. In fact,
conformal geometry enjoys a natural and remarkably elegant formulation as
curved \textsl{Cartan geometry}, and essentially relies on the so-called
Weyl-covariant differential calculus, also known as \textsl{tractor calculus}%
. This is the conformal-covariant generalization of the ordinary
differential calculus; it was originally constructed in \cite{tractor} (%
\textit{cfr.} also \cite{Gover,Curry} for more physicists' minded
treatments, and \cite{RodGover:2012ib} for an application to theAdS/CFT correspondence) and subsequently generalized to all parabolic geometries in \cite{tractorparabolic}.

Minkowski $D$-dimensional space-time $\mathbf{M}^{D-1,1}$ (or the
aforementioned generalizations $\mathbf{M}^{s,D-s}$ thereof) cannot support
a linear implementation of conformal symmetry, and a compactification
procedure, which amounts to adding suitable points at infinity, is needed.
This framework has been formalized and developed by Fefferman and Graham in
\cite{FG}, especially for curved manifolds. A simple instance of flat
conformal geometry is provided by the \textsl{Dirac cone} construction, in
which the $D$-dimensional \textsl{compactified} Minkowski space $\overline{%
\mathbf{M}}^{D-1,1}$ is obtained as a particular section of the space of
light like rays in the so-called \textsl{conformal space} $\mathbf{M}^{D,2}$.

In \cite{Kuzenko}, the compactified $3$-dimensional Minkowski space $%
\overline{\mathbf{M}}^{2,1}$ was constructed, along with its $\mathcal{N}=1$
supersymmetric extension $\mathbf{M}^{2,1|1}$, in terms of a \textsl{%
Lagrangian manifold} over the twistor space $\mathbb{R}^{4}$, by exploiting
the Lie group isomorphism $\mathrm{Spin}(3,2)$ $\cong $ $\mathrm{Sp}(4,%
\mathbb{R})$. Taking inspiration from the isomorphisms (\ref{iso-Lie-1})-(1.5) and also relying on \cite{MS-2-Groups}, in \cite{FL-1} a
symplectic characterization of the $4$-dimensional (compactified and real)
Minkowski space $\overline{\mathbf{M}}^{3,1}$ and $\mathcal{N}=1$ Poincar%
\'{e} superspace $\mathbf{M}^{3,1|1}$ was given, exploiting the Lie group
isomorphism $\mathrm{Spin}(4,2)$ $\cong $ $\widetilde{\mathrm{Sp}}(4,\mathbb{%
C})$. Therein, it was also argued the possibility to extend the approach
also to the other critical dimensions $D=6$ and $10$, thus providing a
uniform and elegant description of $\mathcal{N}=1$ Poincar\'{e} superspaces $%
\mathbf{M}^{q+1,1|1} $ in critical dimensions $D=q+2$ in terms of the four
normed division algebras $\mathbb{A}$'s.\bigskip

In the present paper, we shall be interested in space-time signatures
characterized by the same number of spacelike and timelike dimensions : $s=t$%
. The corresponding signature is usually named \textsl{Kleinian} (or also
\textsl{ultrahyperbolic}). Usually, Susy, SYM's and supergravity theories in
such a signature are investigated by focussing on suitably Wick-rotated
versions of the corresponding theories in Lorentz signature (\textit{cfr.
e.g.} \cite{Klemm-Nozawa}, and Refs. therein). However, also other, more
exotic, possibilities can be considered, such as compactifications of the
so-called ${M}^{\prime }$\textsl{-theory} or ${M}^{\ast }$\textsl{-theory}
(see \textit{e.g.} \cite{Hull-1,Hull-2,Ferrara-Spinors}). Geometries in
Kleinian signature currently remains a vast and yet unexplored realm,
displaying a rich mathematical structure, whose little knowledge is
essentially based on a few studies scattered in literature (\textit{cfr. e.g.%
} \cite{Bryant-1,Dun-1,Dun-2,Dun-3,Hervik,Klemm-Nozawa}).

Although considering Kleinian signature might seem at first a purely
mathematical \textsl{divertissement}, important motivations are actually
provided by Physics. The computation and the study of symmetries of
scattering amplitudes in SYM's and in supergravity highlighted the relevance
of Kleinian signature, especially in $4$ dimensions; indeed, in \cite{OV}
Ooguri and Vafa showed that $D=4$ is the critical dimension of the $\mathcal{%
N}=2$ superstring, whose bosonic part is given by a self-dual metric of
signature $s=t=2$. It is also worth pointing out here that $4$-dimensional
Kleinian signature is essentially related to \textsl{twistors} \cite{Penrose}%
, thus providing a powerful computational tool in the investigation of
scattering amplitudes \cite{Witten}.

The present paper is then devoted to the study of the $4$-dimensional \textsl{Klein space} $\mathbf{M}^{2,2}$, viewed inside the related  \textsl{Klein-conformal} space \footnote{Technically this is called the big cell inside the Klein-conformal (super)space.}, as well as of their supersymmetric extensions, namely
the \textsl{Klein} $\mathcal{N}=1$ \textsl{superspace} $\mathbf{M}^{2,2|1}$
and the corresponding \textsl{Klein-conformal} $\mathcal{N}=1$ \textsl{%
superspace}. By recalling the \textsl{split}
counterparts of the division algebras, namely $\mathbb{A}_{s}$ $=\mathbb{C}%
_{s}$ (\textsl{split complex numbers}), $\mathbb{H}_{s}$ (\textsl{split
quaternions}) and $\mathbb{O}_{s}$ (\textsl{split octonions}), we rely on
the observation that the entries of second row of the order-$2$ \textsl{%
doubly-split} magic square $\mathcal{L}_{2}\left( \mathbb{A}_{s},\mathbb{B}%
_{s}\right) $ \cite{MS,MS2,BS-2}\vspace{1mm}%
\begin{equation}
\begin{tabular}{ccccc}
\hline
&  &  &  &  \\[-4mm]
& $\mathbb{R}$ & $\mathbb{C}_{s}$ & $\mathbb{H}_{s}$ & $\mathbb{O}_{s}$ \\%
[0.1mm]\hline
&  &  &  &  \\[-3mm]
$\mathbb{C}_{s}$ & $\mathfrak{so}(2,1)$ & $\mathfrak{so}(2,2)$ & $\mathfrak{%
so}(3,3)$ & $\mathfrak{so}(5,5)$%
\end{tabular}%
\end{equation}%
can be naturally represented as $\mathfrak{sl}(2,\mathbb{A}_{s})$, then
yielding the isomorphisms of Lie algebras (\textit{cfr. e.g.} \cite{Rios},
and Refs. therein)%
\begin{equation}
\mathfrak{sl}(2,\mathbb{A}_{s})\cong \mathfrak{so}(q/2+1,q/2+1),
\label{iso-Lie-3}
\end{equation}%
where $q$ is here defined as
\begin{equation}
q:=\text{dim}_{\mathbb{R}}\mathbb{A}_{s}=2,4,8\text{~for~}\mathbb{A}_{s}=%
\mathbb{C}_{s},\mathbb{H}_{s},\mathbb{O}_{s}\text{,~respectively},
\label{q-def-2}
\end{equation}%
and $\mathfrak{so}(q/2+1,q/2+1)$ is the Lie algebra of the \textsl{Klein
group} in $D=q+2$. It is then natural to think, in analogy with the non
split case, that the third line of $\mathcal{L}_{2}\left( \mathbb{A}_{s},%
\mathbb{B}_{s}\right) $, \textit{i.e.}%
\begin{equation}
\begin{tabular}{ccccc}
\hline
&  &  &  &  \\[-4mm]
& $\mathbb{R}$ & $\mathbb{C}_{s}$ & $\mathbb{H}_{s}$ & $\mathbb{O}_{s}$ \\%
[0.1mm]\hline
&  &  &  &  \\[-3mm]
$\mathbb{H}_{s}$ & $\mathfrak{so}(3,2)$ & $\mathfrak{so}(3,3)$ & $\mathfrak{%
so}(4,4)$ & $\mathfrak{so}(6,6)$%
\end{tabular}%
\end{equation}%
can be reinterpreted by means of the following Lie algebraic isomorphism
\begin{equation}
\widetilde{\mathfrak{sp}}(4,\mathbb{A}_{s})\cong \mathfrak{so}(q/2+2,q/2+2),
\label{iso-Lie-4}
\end{equation}%
and $\mathfrak{so}(q/2+2,q/2+2)$ is the Lie algebra of the \textsl{%
Klein-conformal group} in $D=q+2$. More in detail, in this paper we give an
explicit proof and take advantage of the Lie group isomorphism $\mathrm{Spin}%
(2,2)\cong \mathrm{SL}(2,\mathbb{H}_{s})$ and $\mathrm{Spin}(3,3)\cong
\mathrm{Sp}(4,\mathbb{C}_{s})$, by constructions similar to the ones made in
\cite{MS-2-Groups} and \cite{FL-1}. While in our treatment the construction
and the Lie group isomorphisms analogues of (\ref{iso-Lie-3}) and (\ref%
{iso-Lie-4}) are explicitly worked out in those cases, nothing\footnote{%
While the generalization to the $D=(3,3)$ is straightforward, the case $%
D=(5,5)$ may be plagued by further issues, which actually arise also in the
Lorentzian case $D=(9,1)$, due to the known problem of constructing the
superconformal algebra in $D>6$ \cite{Nahm}. We aim at tackling this problem
in a future project.} seemingly prevents us from putting forward the
conjecture that our approach equally works well in the other critical
dimensions with ultrahyperbolic signature, \textit{i.e.} in $D=(3,3)$ and in
$D=(5,5)$.

We will point out that the Klein-conformal space in $D=4$, $6$ or $10$
dimensions may be respectively regarded as a certain Lagrangian manifold
over the three aforementioned normed split algebras $\mathbb{A}_{s}$'s. In
fact, the inner motivation of the present analysis also relies on the belief
that a deeper understanding of the relation between Susy and split normed
algebras $\mathbb{A}_{s}$'s from a supergeometric point of view could
provide interesting insights on the classical and quantum properties of
SYM's and supergravity theories in critical dimensions with ultrahyperbolic
signature.

Our approach to $\mathbf{M}^{2,2}$ and its $%
\mathcal{N}=1$ super-extensions will follow closely the one of \cite{FL-1},
which in turn developed a procedure exploited in \cite{flmink,flv}, in which
the \textsl{complex} $4$-dimensional Minkowski (super)space was realized
inside a complex flag (super)manifold, with the
conformal group $\mathrm{SL}(4,\mathbb{C})$ acts naturally. It is here worth
remarking that this is a more physics-oriented approach, in which
superspaces come along with the supergroups describing their
supersymmetries; this is to be contrasted to the approach \textit{e.g.} of
\cite{ma2}, in which super-Grassmannians and superflags are essentially
conceived as complex entities and constructed by themselves. It should also
be recalled that in \cite{flmink,flv} \textsl{real} forms of
four-dimensional Minkowski and conformal (super)spaces were introduced
through suitable involutions, compatible with the natural (supersymmetric)
action of the Poincar\'{e} and conformal $\mathcal{N}=1$, $D=(3,1)$
supergroups.

In the present study, by essentially adapting the treatment of \cite{FL-1}
to Kleinian signature and thus leaving the complex structure and superflags
on the background, we will find a much richer mathematical structure with
respect to the Minkowski case studied in \cite{FL-1} itself. Such a deep
difference can ultimately be traced back to the fact that the action of the
Klein and Klein-conformal group on its irreducible spinor
representation, that can be identified with $\mathbb{C}_{s}^{2}$ and $\mathbb{H}_{s}^{2}$, is \textsl{
not} transitive, and the corresponding spinor space gets then \textsl{stratified}
into \textsl{orbits}, defined by suitable invariant constraints. Remarkably, this has deep consequences in the construction of the Klein (super)space, since one must from the beginning choose a particular pair of orbit representative; in this paper, we focus only on one particular choice of pair of spinors, called \textsl{generic}.
We point out that such a phenomenon of \textsl{spinor stratification} is
absent in Lorentzian signature, in which case the whole spinor
representation space - apart from its origin - consists of a \textsl{unique}
orbit of the (spin covering of the) Lorentz group $\Spin(q+1,1)$. This
uniquely determines the construction of Minkowski and conformal superspaces
\cite{FL-1}. We will determine the isotropy groups (also named \textsl{%
stabilizers}) of the spinor orbits, as well as the constraints which define
them. Relying on the theory of Clifford algebras, spinor algebras and their
representations, we will highlight the relevance of the interplay between
split algebras and the dimensions and reality properties of spinors of
space-time symmetries in Kleinian signature, which in turn are ultimately
based on the representability of the relevant spinor representation spaces
as $2$-dimensional vector spaces over $\mathbb{A}_{s}$ \cite{Sudbery} (also
\textit{cfr.} \cite{Kugo-Townsend}, and Refs. therein).

It is also worth anticipating here that the symmetry of the order-$2$
doubly-split magic square $\mathcal{L}_{2}\left( \mathbb{A}_{s},\mathbb{B}%
_{s}\right) $ (as opposed to the\ order-$2$ split magic square $\mathcal{L}%
_{2}\left( \mathbb{A}_{s},\mathbb{B}\right) $, which is \textsl{not}
symmetric) - promoted to the Lie group level by relying on the work of Dray,
Manogue and collaborators \cite{MS-2-Groups,Dray-2,Dray-3,Dray-4} - will
play an important role in our treatment. Indeed, the Klein-conformal group $%
\mathrm{Spin}(3,3)$ in $4$ dimensions, besides occurring in the entry $%
\mathcal{L}_{2}\left( \mathbb{H}_{s},\mathbb{C}_{s}\right) $ and thus being
characterized as $\mathrm{Spin}(3,3)\cong \widetilde{\mathrm{Sp}}(4,\mathbb{C%
}_{s})$, also appears in the entry $\mathcal{L}_{2}\left( \mathbb{C}_{s},%
\mathbb{H}_{s}\right) $, and as such it enjoys the isomorphism $\mathrm{Spin}%
(3,3)\cong \mathrm{SL}(2,\mathbb{H}_{s})$, as well. In other words, $\mathrm{%
Spin}(3,3)$ can be regarded as the Klein-conformal group in $D=(2,2)$,
namely as $\mathrm{Spin}(q/2+2,q/2+2)$ with $q=2$, or as the Klein group in $%
D=(3,3)$, namely as $\mathrm{Spin}(q/2+1,q/2+1)$ with $q=4$. Since the
spinor stratification of $\mathrm{Spin}(q/2+1,q/2+1)$ over $\mathbb{A}%
_{s}^{2}$ is known, this latter observation immediately allows for the
knowledge of the spinor stratification of the twistor space $\mathbb{C}%
_{s}^{4}\cong \mathbb{H}_{s}^{2}$ relevant for the explicit construction of
the Klein space $\mathbf{M}^{2,2}$ as a suitable
section of the $D=(3,3)$ Klein-conformal space. In our
treatment, we will present an explicit derivation of the aforementioned Lie
group isomorphisms, as well as of the above geometric construction.\medskip

We conclude by briefly mentioning the possible implications of our analysis
for the fascinating task of \textsl{space-time quantization}, on which many
approaches have been pursued and many research venues have been explored in
literature. \textit{E.g.}, in \cite{cfln,cfl,cfl2} the quantum deformation
of the complex (chiral) Minkowski and conformal superspaces was investigated
by exploiting the formal machinery of flag varieties developed in \cite%
{fquant,fquant2}. The more direct approach which stems from the present
study is essentially the one developed in \cite{FL-1}; it exhibits an
intrinsic elegance based on split algebras $\mathbb{A}_{s}$'s, and it may
pave the way to the intriguing task to construct a quantum deformation of
both real Klein and Klein-conformal $\mathcal{N}=1$ superspaces.\bigskip

The plan of the paper is as follows \\

In Section 2 we introduce split composition algebras $\mathbb{A}_{s}$,
setting the notation used in the present work, while in Section 3 we discuss
the construction of quadratic Jordan algebras over $\mathbb{A}_{s}$.

Section
4 reports on the classification of the spinor bundles in critical
dimensions, stressing out the differences between Lorentz and Kleinian
signature.

In Section 5, we focus our attention on the $D=(2,2)$ case, which is related
to the split complex algebra $\mathbb{C}_{s}$, by realizing explicitly the
action of the \textsl{Klein group} on vectors, $2\times 2$ Hermitian
matrices over $\mathbb{C}_{s}$, and spinors, identified with vectors in $%
\mathbb{C}_{s}^{2}$; in particular, we compute the orbit stratification of
spinors, and derive corresponding representatives.

In Section 6, we then extend our analysis to the conformal case, and discuss
the symplectic realization of $\mathrm{Spin}(3,3)$, whose proof can be found
in the Appendix A.

Finally, Section 7 deals with the $D=(2,2)$ construction of the $\mathcal{N}%
=1$ Klein superspace viewed inside the Klein-conformal $%
\mathcal{N}=1$ superspace. In the Appendix B, we also give a short introduction to the basic Supergeometry ingredients needed for a better understanding of this last Section.

\section{Split Algebras}

Addressing the reader to extended treatments given \textit{e.g.} in \cite%
{hypercomplex} and \cite{split-H} (also \textit{cfr.} App. A of \cite{Gun-2}%
, and Refs. therein), we present here some basic definitions on the \textit{%
split algebras} $\mathbb{C}_{s}$ and $\mathbb{H}_{s}$, useful for the
subsequent treatment.\medskip

For each of the composition, normed \textsl{division} algebras $\mathbb{C}$
(complex numbers), $\mathbb{H}$ (Hamilton numbers, or quaternions) and $%
\mathbb{O}$ (Cayley numbers, or octonions), one can respectively construct,
by suitably adapting the Cayley-Dickson procedure, the corresponding \textit{%
split} (composition) algebras $\mathbb{C}_{s}$ (split complex numbers), $%
\mathbb{H}_{s}$ (split quaternions) and $\mathbb{O}_{s}$ (split octonions);
these are characterized by the fact that some of the imaginary units square
to $1$ instead of $-1$.\smallskip

More in detail, one starts constructing the \textit{split complex numbers} $%
\mathbb{C}_{s}$, also named \textit{hyperbolic numbers}, as%
\begin{equation}
\mathbb{C}_{s}:=\{\alpha+j\beta\,|\,j^{2}=1\,\,\,\alpha,\beta\in \mathbb{R}%
\}\,;
\end{equation}%
this algebra is equipped with a natural conjugation
\begin{equation}
a=\alpha+j\beta\longrightarrow \alpha-j\beta=:\overline{a},
\label{conjug-Cs}
\end{equation}%
which is used in order to define the norm%
\begin{equation}
|a|^{2}:=a\overline{a}=\alpha^{2}-\beta^{2}.  \label{norm-Cs}
\end{equation}

Not all elements in $\mathbb{C}_{s}$ are \textsl{invertible}; in fact, it
holds that
\begin{equation}
\frac{1}{a}=\frac{\overline{a}}{|a|^{2}};
\end{equation}%
therefore, an element of $\mathbb{C}_{s}$ with vanishing norm, \textit{i.e.}
$a=\alpha\pm j\alpha$, is \textit{non-invertible}. Then, we denote by $%
\mathbb{C}_{s}^{\times }$ the invertible elements of $\mathbb{C}_{s}$:
\begin{equation}
\mathbb{C}_{s}^{\times }:=\{\alpha+j\beta\,|\,\alpha\neq \pm \beta\}\,.
\label{Cs-inv}
\end{equation}

Every (non-zero) non-invertible element must be of the form $\alpha \mathcal{%
E}$ or $\alpha \overline{\mathcal{E}}$, with $\mathcal{E}:=1+j$ and $\alpha
\in \mathbb{R}$. Moreover, it is here worth noting the following useful
relations:%
\begin{eqnarray}
\mathcal{E}^{2} &=&2\mathcal{E},~~\overline{\mathcal{E}}^{2}=2\overline{%
\mathcal{E}}; \\
\mathcal{E}\overline{\mathcal{E}} &=&0;  \label{EE} \\
a\mathcal{E} &=&(\alpha +\beta )\mathcal{E}\,,\,\,\,\forall \,a=\alpha
+j\beta \in \mathbb{C}_{s}.  \label{res}
\end{eqnarray}%
Moreover, we observe that every element $a=\alpha +j\beta $ can be uniquely
decomposed according to the following
\begin{equation}
a=\alpha _{+}\mathcal{E}+\alpha _{-}\overline{\mathcal{E}}%
\,,\,\,\,\,\,\,\,\alpha _{\pm }:=\frac{1}{2}(\alpha \pm \beta )
\label{decomposition}
\end{equation}%
It should also be remarked that a non-invertible element is always a \textit{%
zero divisor}, due to (\ref{EE}).\smallskip \smallskip

By the iterating the Cayley-Dickson procedure, we then proceed constructing
the \textit{split quaternions}
\begin{equation}
\mathbb{H}_{s}:=\{a+kc\,|\,k^{2}=-1\,\,\,a,c\in \mathbb{C}_{s}\}\,,
\end{equation}%
which, as their divisional counterparts $\mathbb{H}$, are \textsl{%
non-commutative}. Explicitly, any element $h\in \mathbb{H}_{s}$ can be
written as
\begin{equation*}
h =(\underbrace{\alpha+j\beta}_{h_{R}})+k(\underbrace{\gamma+j\delta}%
_{h_{I}})=\alpha+j\,\beta+k\,\gamma+(kj)\,\delta\,, 
\end{equation*}
where $h_{R}$ and $h_{I}$ respectively denote the real and imaginary part of
the split quaternion $h$. Moreover, $j$, $k$ and $kj$ are three
\textquotedblleft imaginary" units, whose multiplication rules are
summarized in the following table :
\begin{equation}
\begin{array}{c|ccc}
& k & kj & j \\ \hline
k & -1 & -j & kj \\
kj & j & 1 & k \\
j & -kj & -k & 1%
\end{array}%
\end{equation}

In $\mathbb{H}_{s}$, the conjugation is defined as%
\begin{equation}
h=h_{R}+kh_{I}\longrightarrow \overline{h}_{R}-kh_{I}=:h^{\ast },
\label{conjug-Hs}
\end{equation}%
or explicitly :
\begin{equation}
h=\alpha+j\beta+k\,\gamma+(kj)\,\delta\longrightarrow
\alpha-j\,\beta-k\,\gamma-(kj)\,\delta=:h^{\ast }.
\end{equation}%
The norm of a split quaternion then reads
\begin{equation}
|h|^{2}:=hh^{\ast }=\alpha^{2}+\gamma^{2}-\beta^{2}-\delta^{2}\,.
\label{norm-Hs}
\end{equation}

It is straightforward to check that the \textit{invertible} split
quaternions $\mathbb{H}_{s}^{\times }$ are given by
\begin{equation}
\mathbb{H}_{s}^{\times
}:=\{\alpha+j\,\beta+k\,(\gamma+j\delta)\,|\,\alpha^{2}+\gamma^{2}\neq
\beta^{2}+\delta^{2}\}.
\end{equation}%
Due to the aforementioned non-commutativity, one should properly discuss
left and right invertibility; nevertheless, it can be proved that left and
right inverse coincide.

It is also worth pointing out that one can construct the following
isomorphism between $\mathbb{H}_{s}$ and the space of $2\times 2$ matrices
with $\mathbb{C}_{s}$-valued entries
\begin{eqnarray}
N &:&=\{M\in \mathbb{M}_{2}(\mathbb{C}_{s})\,|\,\overline{M}\epsilon
=\epsilon M\}\,,  \label{def-N} \\
\epsilon &:&=%
\begin{pmatrix}
0 & 1 \\
-1 & 0%
\end{pmatrix}%
,  \label{epsilon}
\end{eqnarray}%
by means of the map
\begin{equation}
\begin{array}{rrrcl}
Z: & \mathbb{H}_{s} & \rightarrow & N, &  \\[3mm]
& h & \mapsto &
\begin{pmatrix}
h_{R} & h_{I} \\
-\overline{h}_{I} & \overline{h}_{R}%
\end{pmatrix}%
. &
\end{array}
\label{transfo}
\end{equation}%
When considering matrices with $\mathbb{H}_{s}$-valued entries, one can
apply the map $Z$ (\ref{transfo}) entry-wise.\smallskip

Finally, \textit{split octonions} $\mathbb{O}_{s}$ are obtained from $%
\mathbb{H}_{s}$ by further iterating the Cayley-Dickson procedure:
\begin{equation}
\mathbb{O}_{s}:=\{h+lf\,|\,l^{2}=-1\,\,\,h,f\in \mathbb{H}_{s}\}\,.
\end{equation}%
We will not further deal with the algebra $\mathbb{O}_{s}$, since this not
relevant for the present investigation (for a very recent excellent account,
we address to the monography \cite{Tray-Manogue-Book}).\medskip

For convenience in the subsequent treatment, it is here worth recalling the
definition of two symmetries which can be associated to split algebras : the
\textit{norm-preserving} symmetry and the \textit{triality} symmetry.

As it can be seen from (\ref{norm-Cs}) and (\ref%
{norm-Hs}), the squared norm 
of a split algebra element 
is given by the symmetric bilinear form $\eta _{ab}=\eta ^{ab}$ with
signature $\left( \frac{q}{2},\frac{q}{2}\right) $, and $a,b=1,...,q$, with $%
q$ defined in (\ref{q-def-2}) being the real dimension of the split algebra.%
This is in fact the canonical inner product on the Klein space $\mathbf{M}%
^{q/2,q/2}\cong \mathbb{R}^{q/2,q/2}$, which is preserved by $\mathrm{SO}%
(q/2,q/2)=:\mathrm{SO}(\mathbb{A}_{s})$ (whose Lie algebra we denote by $%
\mathfrak{so}\left( q/2,q/2\right) =:\mathfrak{so}(\mathbb{A}_{s})$). Thus, $%
\mathrm{SO}(\mathbb{A}_{s})$ is named as the \textit{norm-preserving group}
of $\mathbb{A}_{s}$ itself.

Then, let us consider the following Lie algebra \cite{Springer-book}:%
\begin{equation}
\mathfrak{tri}(\mathbb{A}_{s}):=\left\{ \left( A,B,C\right) |A\left(
x,y\right) =B(x)y+xC(y),~A,B,C\in \mathfrak{so}\left( q/2,q/2\right)
,~x,y\in \mathbb{A}_{s}\right\} .  \label{triality-def}
\end{equation}%
This algebra, appearing explicitly in the magic square formula of Barton and
Sudbery \cite{BS-1,BS-2} (see also \textit{e.g.} \cite{Evans}), is named as
the \textit{triality symmetry algebra} of $\mathbb{A}_{s}$, and the
corresponding Lie group $Tri\left( \mathbb{A}_{s}\right) $ is referred to as
the \textit{triality group} of $\mathbb{A}_{s}$ itself.

In general, it holds that $\mathrm{SO}(\mathbb{A}_{s})$ is a (not
necessarily proper) subgroup of $Tri\left( \mathbb{A}_{s}\right) $, and thus
one can define the following (symmetric) cosets\footnote{$Id$ denotes the
group identity element throughout.} (for further elucidation, see \textit{%
e.g.} \cite{Gun-2,CFMZ1-D=5,Magic-Coset-Decomp,ADFMT-1}, and Refs. therein) :%
\begin{equation}
\widetilde{\mathcal{A}}_{q}:=\frac{Tri\left( \mathbb{A}_{s}\right) }{\mathrm{%
SO}(\mathbb{A}_{s})}\cong \left\{
\begin{array}{l}
q=2:\mathrm{SO}(1,1), \\
q=4:\mathrm{Sp}(2,\mathbb{R}) \\
q=8:Id,%
\end{array}%
\right.  \label{A-tilde}
\end{equation}%
whose relevance will be exploited further below. For completeness, and later
convenience, we also report the analogue result for the four normed \textsl{%
division} algebras \cite{Hurwitz} $\mathbb{A}=\mathbb{R},\mathbb{C},\mathbb{H%
},\mathbb{O}$ (for which $q=1,2,4,8$, respectively):%
\begin{equation}
\mathcal{A}_{q}:=\frac{Tri\left( \mathbb{A}\right) }{\mathrm{SO}(\mathbb{A})}%
\cong \left\{
\begin{array}{l}
q=1:Id \\
q=2:\mathrm{U}(1), \\
q=4:\mathrm{USp}(2) \\
q=8:Id.%
\end{array}%
\right.  \label{A}
\end{equation}

\section{Quadratic Jordan Algebras over Split Algebras}

Referring to thorough treatments given \textit{e.g.} in \cite{McCrimmon,
Iordanescu} for references and details, we shall here give a brief account
of quadratic Jordan algebras.\medskip

A \textit{Jordan algebra} over a field $\mathbb{F}$ (which we shall
henceforth assume to be $\mathbb{R}$, unless otherwise specified) is an
algebra $J$ with a symmetric product $\circ $%
\begin{equation}
X\circ Y=Y\circ X\in J,~\forall X,Y\in J
\end{equation}%
which satisfies the \textit{Jordan identity}%
\begin{equation}
X\circ (Y\circ X^{2})=(X\circ Y)\circ X^{2},
\end{equation}%
where $X^{2}:=X\circ X$. Therefore, a Jordan algebra is commutative and
generally non-associative.

Given a Jordan algebra $J$, one can define a \textit{norm }$\mathbf{N}$ $%
:J\rightarrow \mathbb{R}$ over it, satisfying the composition property \cite%
{Jacobson}%
\begin{equation}
\mathbf{N}[2X\circ (Y\circ X)-(X\circ X)\circ Y]=\mathbf{N}^{2}(X)\mathbf{N}%
(Y).
\end{equation}%
The \textit{degree} $p$, of the norm form as well as of $J$, is defined by $%
\mathbf{N}(\lambda X)=\lambda ^{p}\mathbf{N}(X)$, where $\lambda \in \mathbb{%
R}$. A \textit{Euclidean} Jordan algebra is a Jordan algebra for which the
condition $X\circ X+Y\circ Y=0$ implies that $X=Y=0$ for all $X,Y\in J$;
they are sometimes called \textsl{compact} Jordan algebras, since their
automorphism groups are compact.\smallskip

In the present investigation, we are interested in a particular class of
simple, \textit{quadratic} Euclidean Jordan algebras (degree $p=2$); the
algebras of such a class \cite{JWVN} are denoted by $J_{2}^{\mathbb{C}_{s}}$%
, $J_{2}^{\mathbb{H}_{s}}$ and $J_{2}^{\mathbb{O}_{s}}$, and they are
generated by Hermitian $(2\times 2)$-matrices over the split composition
algebras $\mathbb{A}_{s}=\mathbb{C}_{s}$, $\mathbb{H}_{s}$, $\mathbb{O}_{s}$%
, respectively :%
\begin{equation}
\mathcal{J}=\left(
\begin{array}{cc}
\alpha & Z \\
\overline{Z} & \beta%
\end{array}%
\right) \in J_{2}^{\mathbb{A}_{s}},  \label{matr}
\end{equation}%
where $\alpha ,\beta \in \mathbb{R}$ and $Z\in \mathbb{A}_{s}$, and the bar
stands for the conjugation pertaining to the algebra under consideration;
moreover, the Jordan product $\circ $ is realized as (one half) the matrix
anticommutator.

The set of linear invertible transformations leaving the quadratic norm of $%
J_{2}^{\mathbb{A}_{s}}$%
\begin{equation}
\mathbf{N}(\mathcal{J}):=\text{det}(\mathcal{J}),~\mathcal{J}\in J_{2}^{%
\mathbb{A}_{s}},  \label{norm-J2}
\end{equation}%
invariant is the so-called \textit{reduced structure group} $Str_{0}\left(
J_{2}^{\mathbb{A}_{s}}\right) $ of $J_{2}^{\mathbb{A}_{s}}$ itself, and it
holds that (recall (\ref{q-def-2}))%
\begin{equation}
Str_{0}\left( J_{2}^{\mathbb{A}_{s}}\right) =\mathrm{Spin}(q/2+1,q/2+1).
\end{equation}%
In other words, the reduced structure group of $J_{2}^{\mathbb{A}_{s}}$ is
the \textit{Klein} \textit{group} $Spin(q/2+1,q/2+1)$ in\footnote{%
Note that $D=q+2$ corresponds to the \textit{critical} space-time dimensions
of superstring theory. In fact, there is a deep relationship between
supersymmetry and division algebras; \textit{cfr. e.g.} \cite%
{Baez-Huerta-1,Huerta-2,Huerta-3,Huerta}, and Refs. therein.} $D=q+2$.

\section{\label{Sec-Spinors}Spinors}

In this Section, we provide some basic definitions and results on spinors,
useful for the subsequent treatment; for further details and elucidation, we
address the reader \textit{e.g.} to \cite{Budinich-1,Budinich-2,Charlton-Th}%
, and Refs. therein.

We will henceforth assume $D=s+t$ \textsl{even} (in view of the specific
case we will be interested in below, namely $D=4$ and $s=t=2$).\medskip

Let us start and consider the properties of (irreducible) spinor
representations of the spin covering group $\mathrm{Spin}(s,t)$ of
pseudo-orthogonal groups $\mathrm{SO}(s,t)$. For more details, \textit{cfr.}
\textit{e.g.} \cite{Spinor-Algebras,Ferrara-Spinors}, and Refs. therein. Let $V$ be a real vector space of dimension $%
D=s+t$, with basis $\left\{ \mathbf{e}_{a}\right\} $ ($a=1,...,D$) and
signature $\left( s,t\right) $ : $V\cong \mathbb{R}^{s,t}$. Then, $V$ admits
a non-degenerate symmetric bilinear form $\eta $ with signature $\left(
s,t\right) $, which in the basis $\left\{ \mathbf{e}_{a}\right\} $ is given
by the metric
\begin{equation}
\eta _{ab}=\eta ^{ab}=\left( \underset{s}{\underbrace{+,...,+}},\underset{t}{%
\underbrace{-,...,-}}\right) .  \label{metric}
\end{equation}

The group $\mathrm{Spin}(V)$ is defined as the unique double-covering of the
identity-connected component of $\mathrm{SO}(s,t)$. A spinor representation
of $\mathrm{Spin}(V)^{\mathbb{C}}$ is an irreducible complex representation
whose highest weights are the fundamental weights corresponding - within
usual convention - to the right extreme nodes in the Dynkin diagram.

A spinor representation of $\mathrm{Spin}(V)$ over the reals $\mathbb{R}$
(which we will be interested in) is an irreducible representation over $%
\mathbb{R}$, whose complexification is a direct sum of spin representations.
Two parameters, namely the signature $\rho :=s-t$ mod$(8)$ and the dimension
$D=s+t$ mod$(8)$, classify the properties of the spinor representation (%
\textit{cfr. e.g.} \cite{Spinor-Algebras}, and Refs. therein).

When $s=t$ (and thus $\rho =0$), the real space $V\cong \mathbb{R}^{s,s}$ is
named \textit{Klein space}, its signature $\left( s,t\right) =(s,s)$ \textit{%
Kleinian} (or \textit{hyperbolic}), and the corresponding spin group $%
\mathrm{Spin}(s,s)$ is named \textit{Klein group}.

\subsection{Pure Spinors}

The \textit{Clifford algebra}\footnote{%
Note that in general $\mathcal{C}(s,t)$ is \textsl{not} isomorphic to $%
\mathcal{C}(t,s)$, even if $\mathrm{Spin}(s,t)\cong \mathrm{Spin}(t,s)$ (and
thus $\mathrm{SO}(s,t)\cong \mathrm{SO}(t,s)$); \textit{cfr. e.g.} \cite%
{Spinor-Algebras,flmink}.} $\mathcal{C}(s,t)$ associated to $V$ is generated
by the $s+t$ Dirac gamma matrices $\Gamma ^{a}$'s obeying%
\begin{equation}
\left\{ \Gamma ^{a},\Gamma ^{b}\right\} =2\eta ^{ab}\mathbb{I},
\end{equation}%
where $\mathbb{I}$ denotes the identity matrix. By $\psi $ we denote a $2^{(s+t)/2}$-dimensional spinor,
namely a vector of the $2^{(s+t)/2}$-dimensional representation space $S$ of
$\mathcal{C}(s,t)$; for $z\in V$, $\psi $ is defined by the Cartan equation
\cite{Cartan}%
\begin{equation}
z_{a}\Gamma ^{a}\psi =0,  \label{C-Eq}
\end{equation}%
yielding the existence of a \textit{totally null} plane of dimension $%
d\leqslant (s+t)/2$, denoted by $T_{d}(\psi )$. In $D=s+t$ \textsl{even}
dimensions (as we are assuming throughout; \textit{cfr.} the start of the
present Section), $\psi $ does \textsl{not} provide an \textsl{irreducible}
representation for $\mathrm{Spin}(s,t)$.

A \textsl{\textquotedblleft volume element"} in the Clifford algebra $%
\mathcal{C}(s,t)$ can be defined by introducing the gamma matrix $\Gamma
_{s+t+1}:=\Gamma _{1}\Gamma _{2}...\Gamma _{s+t}$, which anticommutes with
all $\Gamma _{a}$'s; it can be used to construct an invariant projector $%
\mathbb{P}_{\pm}$ and we denote by $\psi ^{\pm }$ the \textit{chiral} (or
\textit{Weyl}) spinors, namely the $2^{(s+t)/2-1}$-dimensional spinors
defined by
\begin{equation}
\psi ^{\pm }:=\mathbb{P}_{\pm} \psi ,
\end{equation}%
implying the corresponding chiral Cartan--Weyl equations to read%
\begin{equation}
z_{a}\Gamma ^{a}\mathbb{P}_{\pm} \psi =0.  \label{CW-Eqs}
\end{equation}%
Eq. (\ref{CW-Eqs}) define a $d$-dimensional totally null plane $T_{d}(\psi
^{\pm })$, and each of the chiral spinors $\psi ^{\pm }$ provides an \textsl{%
irreducible} representation for $\mathrm{Spin}(s,t)$. The existence of
chiral spinors determines the splitting of the $\mathcal{C}(s,t)$%
-representation space $S$ (with generic element $\psi $) into the direct sum
of two $\mathrm{Spin}(s,t)$-representation spaces $S^{\pm }$ (with generic
elements $\psi ^{\pm }$) :

\begin{equation}
S=S^{+}\oplus S^{-}.  \label{chiral-split}
\end{equation}

For $d=(s+t)/2$, \textit{i.e.} for the \textsl{maximal} dimension of $%
T_{d}(\psi ^{\pm })$, the corresponding Weyl spinor $\psi ^{\pm }$ is named
\textit{pure}, and $T_{\left( s+t\right) /2}(\psi ^{\pm })\cong \pm \psi
^{\pm }$ \cite{Cartan}. Cartan himself stressed out the importance of this
equivalence, which indeed establishes the crucial link between spinor
geometry and \textsl{projective} Euclidean geometry. Actually, Cartan named
such spinors \textit{simple}, and the nowadays customary naming \textit{pure}
is due to Chevalley \cite{Chevalley}.

It should be remarked that the dimension of $T_{(s+t)/2}(\psi ^{\pm })$
increases linearly with $(s+t)/2$, while that of the pure $\psi ^{\pm }$'s
increases as $2^{(s+t)/2-1}$; consequently, for high $(s+t)/2$'s, pure
spinors will be given by the solutions of suitable (quadratic) constraining
relations, named \textit{pure spinor constraints}, which allow to separate
(in a $\mathrm{Spin}(V)$-invariant way) the space of pure spinors from the
space of \textsl{\textquotedblleft impure"} ones. In fact, \textit{all}
spinors are pure for $\left( s+t\right) /2=1,2,3$ (\textit{i.e.} in $D=2,4,6$
dimensions), while for $(s+t)/2=4,5,6,7,...$ (\textit{i.e.} in $%
D=8,10,12,14,...$ dimensions ) pure spinors are subject to $1$, $10$, $66$, $%
364$, $...$ constraints, respectively; in general, in $D=s+t$ dimensions
there are $\binom{s+t}{(s+t)/2-4}$ pure spinor constraints.

For instance, in $D=s+t=10$ dimensions, there are $10$ pure spinor
constraints, given by%
\begin{equation}
\psi \Gamma ^{a}\psi =0,~\forall a=1,...,10,  \label{pure-D=10}
\end{equation}%
which are especially relevant for the formulation of the \textit{pure spinor
formalism} of superstrings \cite{Berkovits} (see \textit{e.g.} \cite{PSF}
for an introduction).

\subsection{Classification}

The problem of classifying spinors is usually formulated in subsequent steps
as : \textbf{(i)} determining the structure of the spinor orbits $\mathcal{O}
$'s under the action of the $\mathrm{Spin}$ group; \textbf{(ii)} computing
the isotropy (\textit{stabilizer}) group $\mathcal{H}\subset \mathrm{Spin}$
of each orbit $\mathcal{O}$; and \textbf{(iii)} determining the algebra of
invariants of the spinor representation space $S$.

The \textit{orbit} $\mathcal{O}_{\psi }$ of a well-defined spinor
representative $\psi $ under the $\mathrm{Spin}$ group is a coset manifold,
whose structure is determined by the isotropy group $\mathcal{H}_{\psi }$ of
$\psi $ :%
\begin{equation}
\mathcal{O}_{\psi }\cong \frac{\mathrm{Spin}}{\mathcal{H}_{\psi }};
\end{equation}%
in general, the embedding of $\mathcal{H}_{\psi }$ into $\mathrm{Spin}$ is
not maximal nor symmetric; thus, the coset $\mathcal{O}_{\psi }$ is usually
non-symmetric.\smallskip

Classification of spinors was first studied by Chevalley \cite{Chevalley},
who considered the orbit of pure spinors. He found that, in general, the
orbit of pure spinors is the orbit of \textsl{least} dimension (or,
equivalently, the stabilizer of pure spinors is the \textsl{largest} one
among all spinor stabilizers). Chevalley's analysis classifies spinors in
all dimensions up to $D=s+t=6$; as mentioned above, in these cases \textsl{%
all spinors are pure}.

Igusa has then classified spinors in dimensions up to $D=s+t=12$ \cite{Igusa}%
. For each spinor orbit, he provided a well-defined representative, as well
as the stabilizer of the orbit itself. Using similar techniques, full
classifications of spinors have been worked out in more than $12$ dimensions
by Kac and Vinberg \cite{KV78}, Popov \cite{Pop80}, Zhu \cite{Zhu92},
Antonyan and Elashvili \cite{AE82}, but very little is known beyond $16$
dimensions. A nice summary of the spinor classification programme has been
recently accounted in \cite{Charlton-Th} (for what concerns pure spinors,
see also \textit{e.g.} \cite{Pure-Spinors-Polish,Furlan}).\smallskip

Spinors in \textsl{critical} dimensions $D=s+t=q+2=3,4,6,10$ have also been
studied by Bryant \cite{Bryant-1,Bryant-2}, whose approach exploited the
connection between spinors and the four normed division algebras $\mathbb{A}=%
\mathbb{R},\mathbb{C},\mathbb{H},\mathbb{O}$. As a physical application,
such results have been recently applied to the gauging of $\mathcal{N}=(1,0)$
\textit{magic} \cite{GST} chiral supergravities in $D=6$ (Lorentzian : $%
s=5,t=1$) space-time dimensions in \cite{Gunaydin-D=6}.

\subsection{\label{Lorentz-vs-Klein}Spinors and Space-Time Signature :
Lorentz \textsl{versus} Klein}
\label{spinors}
Before treating in some detail the irreducible spinor representations of the
\textit{Klein group} $\mathrm{Spin}(2,2)$ in Sec. \ref{Spin(2,2)} (which
will then be instrumental for the introduction of the Klein and conformal $%
D=(2,2)$ $\mathcal{N}=1$ superspaces in Sec. \ref{supermink}), we now
briefly recall the crucial differences between Lorentzian and Klein spinors
in critical dimensions $D=q+2$ (for $q=2,4,8$), especially for what concerns
the \textsl{representability} in terms of division and split algebras,
respectively. In the specific case of $D=4$, this reasoning will also
highlight the important differences between the approach exploited in the
present investigation and the one considered in \cite{FL-1} (note that we
will anticipate some results, which will then be obtained and discussed in
the treatment of subsequent Sections).\smallskip

As far as notation is concerned, by $M_{p}(\mathbb{R})$ ($M_{p}(\mathbb{C})$%
) we will denote the algebra of $p\times p$ matrices with entries in the $%
\mathbb{R}$ ($\mathbb{C}$) (consistently with (\ref{def-N})). Instead, $%
M_{p}(\mathbb{H})$ will denote the set of $p\times p$ complex matrices
satisfying the \textsl{quaternionic condition}
\begin{equation}
\overline{M}=-\Omega M\Omega ,  \label{finite-sympl}
\end{equation}%
where the bar denotes conjugation in $\mathbb{C}$, and $\Omega $ is the
symplectic metric (for $p=2$, $\Omega =\epsilon $ (\ref{epsilon})). If $%
\Omega $ is non-degenerate, (\ref{finite-sympl}) implies $p$ to be \textsl{%
even}, and $M$ can be written as a $p/2\times p/2$ matrix whose entries are
quaternionic. It should also be stressed that we will be considering the
Clifford algebras as real algebras throughout (\textit{cfr. e.g.} Tables 1
and 2 of \cite{Spinor-Algebras}).

\begin{itemize}
\item $D=10$ ($\leftrightarrow q=8$, thus corresponding to $\mathbb{O}_{s}$
or $\mathbb{O}$). Let us first consider the \textbf{Klein case} : $D=(5,5)$,
namely $s=t=5$, and thus $\rho =0$. The Clifford algebra $\mathcal{C}(5,5)$,
as a real algebra, is isomorphic to \textsl{real} $32\times 32$ matrices :%
\begin{equation}
\mathcal{C}(5,5)\cong M_{32}(\mathbb{R}),
\end{equation}%
with dim$_{\mathbb{R}}\mathcal{C}(5,5)=32^{2}=2^{10}$. The spinor
representation space $S$ of $\mathcal{C}(5,5)$ is \textsl{real}, with real
dimension $2^{5}=32$, and it splits into chiral spinor representation spaces
$S^{\pm }$ as given by (\ref{chiral-split}). Each of $S^{\pm }$ is \textsl{%
real}, with real dimension $2^{4}=16$ : namely, it is a \textsl{Majorana-Weyl%
} spinor representation space. After\footnote{\cite{Sudbery} only deals with the division case. However, the treatment for
the split case goes through almost without modification. Indeed, it is known
that the $\mathbf{27}$ of $E_{6(6)}$ is $J_{3}\left( \mathbb{O}_{s}\right) $%
, so it is essentially ensured that the $\mathbf{16}$ of $Spin(5,5)$ is
representable by $\mathbb{O}_{s}^{2}$. The same holds, by suitable algebraic
truncations, for $\mathbb{H}_{s}^{2}$ and $\mathbb{C}_{s}^{2}$. We thank
Leron Borsten for correspondence on this.} \cite{Sudbery} (also \textit{cfr.%
} \cite{Kugo-Townsend}, and Refs. therein), a Majorana-Weyl spinor $\psi
^{\pm }$ of $\mathrm{Spin}(5,5)\cong \mathrm{SL}(2,\mathbb{O}_{s})$ can be
represented by a vector in\footnote{%
In the dimension-labelled physicists' notation of the group irreprs., the
dimensions are real, unless otherwise noted by suitable subscripts.} $%
\mathbb{O}_{s}^{2}$ (from (\ref{A-tilde}), recall that $\widetilde{\mathcal{A%
}}_{8}\cong Id$) :%
\begin{equation}
\left.
\begin{array}{c}
\psi ^{+} \\
\psi ^{-}%
\end{array}%
\right\} \cong \mathbb{O}_{s}^{2}\cong \left\{
\begin{array}{c}
\mathbf{16} \\
\mathbf{16}^{\prime }%
\end{array}%
\right. ~\text{of~}\mathrm{Spin}(5,5)\mathbf{.}
\end{equation}%
Let us then consider the \textbf{Lorentz case} : $D=(9,1)$, namely $s=9$, $%
t=1$, and thus $\rho =8=0$ mod$\left( 8\right) $. Since $\rho $ and $D$ are
the same as the Klein case previously considered, the spinor properties
coincide. Indeed, the Clifford algebra $\mathcal{C}(9,1)$, as a real
algebra, is isomorphic to \textsl{real} $32\times 32$ matrices :%
\begin{equation}
\mathcal{C}(9,1)\cong M_{32}(\mathbb{R}),
\end{equation}%
with dim$_{\mathbb{R}}\mathcal{C}(9,1)=32^{2}=2^{10}$, and the spinor
representation space $S$ of $\mathcal{C}(9,1)$ is \textsl{real}, with real
dimension $2^{5}=32$. Each of the chiral spinor representation spaces $%
S^{\pm }$ is \textsl{Majorana-Weyl}, with real dimension $2^{4}=16$. Once
again, after \cite{Sudbery} (also \textit{cfr.} \cite{Kugo-Townsend}, and
Refs. therein), a Majorana-Weyl spinor $\psi ^{\pm }$ of $\mathrm{Spin}%
(9,1)\cong \mathrm{SL}(2,\mathbb{O})$ can be represented by a vector in $%
\mathbb{O}^{2}$ (from (\ref{A}), recall that $\mathcal{A}_{8}\cong Id$) :%
\begin{equation}
\left.
\begin{array}{c}
\psi ^{+} \\
\psi ^{-}%
\end{array}%
\right\} \cong \mathbb{O}^{2}\cong \left\{
\begin{array}{c}
\mathbf{16} \\
\mathbf{16}^{\prime }%
\end{array}%
\right. ~\text{of~}\mathrm{Spin}(9,1)\mathbf{.}
\end{equation}

\item $D=6$ ($\leftrightarrow q=4$, thus corresponding to $\mathbb{H}_{s}$
or $\mathbb{H}$). Let us first consider the \textbf{Klein case} : $D=(3,3)$,
namely $s=t=3$, and thus $\rho =0$. The Clifford algebra $\mathcal{C}(3,3)$,
as a real algebra, is isomorphic to \textsl{real} $8\times 8$ matrices :%
\begin{equation}
\mathcal{C}(3,3)\cong M_{8}(\mathbb{R}),
\end{equation}%
with dim$_{\mathbb{R}}\mathcal{C}(3,3)=8^{2}=2^{6}$. The spinor
representation space $S$ of $\mathcal{C}(3,3)$ is \textsl{real}, with real
dimension $2^{3}=8$, and it splits into chiral spinor representation spaces $%
S^{\pm }$, which are also \textsl{real} and with real dimension $2^{2}=4$ :
namely, they are \textsl{Majorana-Weyl} $4$-dimensional spinor
representation spaces. Therefore, a generic element $\psi =\psi ^{+}\oplus
\psi ^{-}\in S$, namely a \textsl{non-chiral} spinor of $\mathrm{Spin}%
(3,3)\cong \mathrm{SL}(4,\mathbb{R})\cong \mathrm{SL}(2,\mathbb{H}_{s})\cong
\widetilde{\mathrm{Sp}}(4,\mathbb{C}_{s})$ (\textit{cfr.} (\ref{iso-3}) below), can be represented by a
vector in $\mathbb{H}_{s}^{2}$ :%
\begin{equation}
\psi =\psi ^{+}\oplus \psi ^{-}\cong \mathbb{H}_{s}^{2}\cong \left( \mathbf{%
4,2}\right) ~\text{of~}\mathrm{Spin}(3,3)\times \widetilde{\mathcal{A}}_{4}%
\mathbf{,}  \label{(3,3)}
\end{equation}%
where $\widetilde{\mathcal{A}}_{4}\cong \mathrm{SL}(2,\mathbb{R})\cong
\mathrm{Sp}(2,\mathbb{R})$ has been recalled from (\ref{A-tilde}). Note that
the presence of a non-trivial $\widetilde{\mathcal{A}}_{q}\neq Id$ (\ref%
{A-tilde}) is crucial for the consistency of the spinor properties with the
representability in terms of split algebras. Let us then consider the
\textbf{Lorentz case} : $D=(5,1)$, namely $s=5$, $t=1$, and thus $\rho =4$.
The Clifford algebra $\mathcal{C}(5,1)$, as a real algebra, is isomorphic to
\textsl{quaternionic} $4\times 4$ matrices (in the sense specified above) :%
\begin{equation}
\mathcal{C}(5,1)\cong M_{4}(\mathbb{H}),
\end{equation}%
with dim$_{\mathbb{C}}\mathcal{C}(5,1)=4^{2}=2^{4}$. Thus, the spinor
representation space $S$ of $\mathcal{C}(5,1)$ is \textsl{quaternionic},
with \textsl{complex} dimension $2^{3}=8$. Each of the chiral spinor
representation spaces $S^{\pm }$ is \textsl{quaternionic}, with \textsl{%
complex} dimension $2^{2}=4$. After \cite{Sudbery} (also \textit{cfr.} \cite%
{Kugo-Townsend}, and Refs. therein), a \textsl{quaternionic} (also named
\textit{symplectic-Majorana-Weyl}) spinor $\psi ^{\pm }$ of $\mathrm{Spin}%
(5,1)\cong \mathrm{SU}^{\ast }(4)\cong \mathrm{SL}(2,\mathbb{H})$ can be
represented by a vector in $\mathbb{H}^{2}$ :%
\begin{equation}
\left.
\begin{array}{c}
\psi ^{+} \\
\psi ^{-}%
\end{array}%
\right\} \cong \mathbb{H}_{s}^{2}\cong \left\{
\begin{array}{c}
\left( \mathbf{4,2}\right) \\
\left( \overline{\mathbf{4}}\mathbf{,2}\right)%
\end{array}%
\right. ~\text{of~}\mathrm{Spin}(5,1)\times \mathcal{A}_{4}\mathbf{,}
\label{(5,1)}
\end{equation}%
where $\mathcal{A}_{4}\cong \mathrm{SU}(2)\cong \mathrm{USp}(2)$ has been
recalled from (\ref{A}). Again, let us point out that the presence of a
non-trivial $\mathcal{A}_{q}\neq Id$ (\ref{A}) is crucial for the
consistency of the spinor properties with the representability in terms of
division algebras\footnote{%
Concerning physical applications, the relevance of $\mathcal{A}_{q}$ (\ref{A}%
) as a part of the $U$-duality symmetry of $\mathcal{N}=(1,0)$ chiral
\textsl{magic} supergravity theories in $D=(5,1)$ dimensions has been
recently exploited in \cite{Gunaydin-D=6} (\textit{cfr.} Table 2 and Sec.
3.2 therein).}. Note that in (\ref{(5,1)}) the bar denotes the conjugation
in $\mathbb{C}$.

\item $D=4$ ($\leftrightarrow q=2$, thus corresponding to $\mathbb{C}_{s}$
or $\mathbb{C}$). Let us first consider the \textbf{Klein case} : $D=(2,2)$,
namely $s=t=2$, and thus $\rho =0$; this will be the case considered in
detail in the next Sections. The Clifford algebra $\mathcal{C}(2,2)$, as a
real algebra, is isomorphic to \textsl{real} $4\times 4$ matrices :%
\begin{equation}
\mathcal{C}(2,2)\cong M_{4}(\mathbb{R}),  \label{C(2,2)}
\end{equation}%
with dim$_{\mathbb{R}}\mathcal{C}(3,3)=4^{2}=2^{4}$. The spinor
representation space $S$ of $\mathcal{C}(3,3)$ is \textsl{real}, with real
dimension $2^{2}=4$, and it splits into chiral spinor representation spaces $%
S^{\pm }$, which are also \textsl{real} and with real dimension $2$ :
namely, they are \textsl{Majorana-Weyl} $2$-dimensional spinor
representation spaces. Thus, a generic element $\psi =\psi ^{+}\oplus \psi
^{-}\in S$, namely a \textsl{non-chiral} spinor of $\mathrm{Spin}(2,2)\cong
\mathrm{SL}(2,\mathbb{R})\times \mathrm{SL}(2,\mathbb{R})\cong \mathrm{SL}(2,%
\mathbb{C}_{s})$ (\textit{cfr.} (\ref{isso}) below), can be represented by a
vector in $\mathbb{C}_{s}^{2}$ :%
\begin{equation}
\psi =\psi ^{+}\oplus \psi ^{-}\cong \mathbb{C}_{s}^{2}\cong \left( \mathbf{%
2,1}\right) _{+}+\left( \mathbf{1},\mathbf{2}\right) _{-}~\text{of~}\mathrm{%
Spin}(2,2)\times \widetilde{\mathcal{A}}_{2}\mathbf{,}  \label{(2,2)}
\end{equation}%
where the \textquotedblleft $+$" and \textquotedblleft $-$" subscripts
denote weights with respect to $\widetilde{\mathcal{A}}_{2}\cong \mathrm{SO}%
(1,1)$ (\textit{cfr.} (\ref{A-tilde})). Again, we observe that the presence
of a non-trivial $\widetilde{\mathcal{A}}_{q}\neq Id$ (\ref{A-tilde}) is
crucial for the consistency of the spinor properties with the
representability in terms of split algebras. Also, note the \textsl{%
non-simple} nature of $\mathrm{Spin}(2,2)\cong \mathrm{SL}(2,\mathbb{R}%
)\times \mathrm{SL}(2,\mathbb{R})$ yields the spinor split $\psi =\left(
\mathbf{2,1}\right) _{+}+\left( \mathbf{1},\mathbf{2}\right) _{-}$, as well
as the chirality interpretation of $\widetilde{\mathcal{A}}_{2}$ itself (see
below). Let us then consider the \textbf{Lorentz case} : $D=(3,1)$, namely $%
s=3$, $t=1$, and thus $\rho =2$. The Clifford algebra $\mathcal{C}(3,1)$, as
a real algebra, is isomorphic to \textsl{real} $4\times 4$ matrices :%
\begin{equation}
\mathcal{C}(3,1)\cong M_{4}(\mathbb{R}),
\end{equation}%
with dim$_{\mathbb{R}}\mathcal{C}(5,1)=4^{2}=2^{4}$. The spinor
representation space $S$ of $\mathcal{C}(3,1)$ is \textsl{real}, with real
dimension $2^{2}=4$. Each of the chiral spinor representation spaces $S^{\pm
}$ is \textsl{complex}, with \textsl{complex} dimension $2$. Therefore, a
\textsl{chiral complex} spinor $\psi ^{+}$ (or $\psi ^{-}$) of $\mathrm{Spin}%
(3,1)\cong \mathrm{SL}(2,\mathbb{C})$ can be represented by a vector in $%
\mathbb{C}^{2}$ :%
\begin{eqnarray}
\psi ^{+} &\cong &\mathbb{C}^{2}\cong \mathbf{2}_{\mathbb{C}}\text{ of }%
\mathrm{SL}(2,\mathbb{C})\equiv \left( \mathbf{2,1}\right) _{+}+\left(
\mathbf{1},\mathbf{2}\right) _{-}~\text{of~}\mathrm{Spin}(3,1)\times
\mathcal{A}_{2}\mathbf{;}  \label{(3,1)} \\
\psi ^{-} &\cong &\overline{\psi ^{+}}\cong \mathbb{C}^{2}\cong \overline{%
\mathbf{2}}_{\mathbb{C}}\text{ of }\mathrm{SL}(2,\mathbb{C})\equiv \left(
\mathbf{2,1}\right) _{-}+\left( \mathbf{1},\mathbf{2}\right) _{+}~\text{of~}%
\mathrm{Spin}(3,1)\times \mathcal{A}_{2}\mathbf{;}  \label{(3,1)-2}
\end{eqnarray}%
where the \textquotedblleft $+$" and \textquotedblleft $-$" subscripts here
denote charges with respect to $\mathcal{A}_{2}\cong U(1)$ (\textit{cfr.} (%
\ref{A})). Again, we stress that the presence of a non-trivial $\mathcal{A}%
_{q}\neq Id$ (\ref{A}) is crucial for the consistency of the spinor
properties with the representability in terms of division algebras. The
comparison between (\ref{(2,2)}) and (\ref{(3,1)})-(\ref{(3,1)-2}) explains
the necessary differences between the approach exploited in the present
investigation and the one considered in \cite{FL-1}. Note that in (\ref%
{(3,1)-2}) the bar in $\overline{\mathbf{2}}_{\mathbb{C}}$ denotes the
conjugation in $\mathbb{C}$, whereas the bar in $\overline{\psi ^{+}}$
denotes the spinor conjugation, which in turn - because of the
representability $\psi ^{+}\cong \mathbb{C}^{2}$ - is \textsl{induced} by
the conjugation in $\mathbb{C}$ itself. 
\end{itemize}

\section{\label{Spin(2,2)}Vectors and Spinors of the Klein group $\mathrm{%
Spin}(2,2)$}

We are now going to consider in some detail the irreducible spinor
representations of the \textit{Klein group} $\mathrm{Spin}(2,2)$, namely of $%
\mathrm{Spin}(V)$, where $V$ is the \textit{Klein space} $\mathbf{M}%
^{2,2}\cong \mathbb{R}^{2,2}$. As mentioned above, this latter is a $4$%
-dimensional real vector space with Kleinian signature, \textit{i.e.} with $%
s=2$ spacelike dimensions and $t=2$ timelike dimensions (thus, having $\rho
=0$).

As reported in Sec. \ref{Lorentz-vs-Klein}, the theory of spinor algebras
(see \textit{e.g.} \cite{Spinor-Algebras}) yields that the \textsl{non-chiral%
} spinor representation $\psi $ is \textsl{real}, of dimension $2^{D/2}=4$.
This provides an irreducible representation of the Clifford algebra $%
\mathcal{C}(2,2)$ (\ref{C(2,2)}); however, since $D=4$ is \textsl{even},
such a representation $\psi $ is \textit{not} irreducible under $\mathrm{Spin%
}(2,2)$, and the corresponding representation space $S$ splits into two $%
\mathrm{Spin}(2,2)$-irreducible \textsl{Majorana-Weyl} spinor subspaces%
\footnote{%
In this case, the chiral projectors on $S^{\pm }$ are real, as well.}, as
given by (\ref{chiral-split}), each of real dimension $2$. Thus, one can
reconsider (\ref{(2,2)}), writing%
\begin{equation}
\psi_{(2,2)} =\underset{\psi ^{+}}{(\mathbf{2},%
\mathbf{1})_{+}}\oplus \underset{\psi ^{-}}{(\mathbf{1},\mathbf{2})_{-}}%
~\cong \left(
\begin{array}{c}
a \\
c%
\end{array}%
\right) ,~a,c\in \mathbb{C}_{s}.  \label{Weyl}
\end{equation}%
As noted below (\ref%
{(2,2)}), subscripts \textquotedblleft $+$" and \textquotedblleft $-$" in (%
\ref{Weyl}) denote weights with respect to $\widetilde{\mathcal{A}}_{2}\cong
\mathrm{SO}(1,1)$ (\ref{A-tilde}); on the other hand, they also represent
the chirality, since $\psi ^{+}$ and $\psi ^{-}$ are \textsl{Majorana-Weyl}
spinors of real dimension $2$ with \textsl{opposite} chirality. Thus, in $D=4
$ Kleinian dimensions $\widetilde{\mathcal{A}}_{2}\cong \mathrm{SO}(1,1)$,
commuting with the Klein group $\mathrm{Spin}\left( 2,2\right) $, can
actually be identified the chirality operator in $\mathbb{C}_{s}^{2}$.

Summarizing, $\mathrm{Spin}\left( 2,2\right) \times \mathrm{SO}(1,1)$, has
the following three representations of (real) dimension $4$ :

\begin{enumerate}
\item The (\textsl{non-chiral}) \textsl{spinor} representation $\psi $ (\ref%
{Weyl}).

\item Its \textsl{conjugate} \textsl{spinor} representation%
\begin{equation}
\overline{\psi }_{(2,2)} =\underset{\overline{\psi
^{+}}}{(\mathbf{2},\mathbf{1})_{-}}\oplus \underset{\overline{\psi ^{-}}}{(%
\mathbf{1},\mathbf{2})_{+}}~\cong \left(
\begin{array}{c}
\overline{a} \\
\overline{c}%
\end{array}%
\right) ,  \label{Weyl-conjug}
\end{equation}%
where it is immediate to realize that, by virtue of the representability of $%
\psi $ as $\mathbb{C}_{s}^{2}$, the conjugation in $S$ is \textsl{induced}
by the conjugation\footnote{%
After the remarks below (\ref{(3,1)})-(\ref{(3,1)-2}), the same holds in $%
D=(3,1)$, as a consequence of the representability in terms of $\mathbb{C}%
^{2}$.} (\ref{conjug-Cs}) in $\mathbb{C}_{s}$.

\item The \textit{vector} $x:=\left( \mathbf{2},\mathbf{2}\right) _{0}$,
which (differently from the spinor representations at points 1 and 2 above)
\textsl{descends} to an irreducible representation of $\mathrm{SO}%
(2,2)\left( \times \mathrm{SO}(1,1)\right) $. Consistently, it is given by
the tensor product of the Majorana-Weyl spinors $\psi ^{+}$ and $\psi ^{-}$
(or of their conjugate; \textit{cfr. e.g.} Table 3 of \cite{Spinor-Algebras}%
, with $D=4$ and $k=1$):%
\begin{equation}
\underset{\left( \mathbf{2},\mathbf{2}\right) _{0}}{x}~=~\underset{(\mathbf{2%
},\mathbf{1})_{+}}{\psi ^{+}}\otimes \underset{(\mathbf{1},\mathbf{2})_{-}}{%
\psi ^{-}}~=~\underset{(\mathbf{2},\mathbf{1})_{-}}{\overline{\psi ^{+}}}%
\otimes \underset{(\mathbf{1},\mathbf{2})_{+}}{\overline{\psi ^{-}}}.
\label{vector}
\end{equation}%
$x$ (\ref{vector}) can be consistently represented as an element of $J_{2}^{%
\mathbb{C}_{s}}$, as follows. In the standard basis of $\mathbf{M}^{2,2}$, $%
x^{a}=(x^{1},\cdots ,x^{4})$ ($a=1,...,4=s+t$); then, its components can be
rearranged as entries of the following $2\times 2$ Hermitian matrix (recall (%
\ref{matr}) and (\ref{iso-2})):%
\begin{equation}
\mathcal{X}:=%
\begin{pmatrix}
x_{+} & \overline{a} \\
a & x_{-}%
\end{pmatrix}%
\in J_{2}^{\mathbb{C}_{s}},  \label{X-call}
\end{equation}%
where $a:=x^{3}+jx^{2}\in \mathbb{C}_{s}$, and $\mathbb{R}\ni x_{\pm
}:=x^{1}\pm x^{4}$, and the bar denotes the conjugation in $\mathbb{C}_{s}$
(see (\ref{conjug-Cs})). The so-called \textit{trace reversal }$\widetilde{%
\mathcal{X}}$ of $\mathcal{X}$ is defined as follows :%
\begin{equation}
\widetilde{\mathcal{X}}:=-%
\begin{pmatrix}
x_{-} & -\overline{a} \\
-a & x_{+}%
\end{pmatrix}%
\in J_{2}^{\mathbb{C}_{s}}\,.  \label{X-call-tilde}
\end{equation}%
Then, by recalling (\ref{norm-Cs}), we observe that
\end{enumerate}

\begin{equation}
\text{det}\mathcal{X}=\left( x^{1}\right) ^{2}-\left( x^{4}\right)
^{2}-|z|^{2}=\left( x^{1}\right) ^{2}+\left( x^{2}\right) ^{2}-\left(
x^{3}\right) ^{2}-\left( x^{4}\right) ^{2}=\eta _{ab}x^{a}x^{b},
\end{equation}%
or equivalently

\begin{equation}
\mathcal{X}\widetilde{\mathcal{X}}=\widetilde{\mathcal{X}}\mathcal{X}=:-\eta
_{ab}x^{a}x^{b}\,\mathbb{I}\,,
\end{equation}%
where the metric $\eta _{ab}=\eta ^{ab}$ is given by (\ref{metric}) with $%
s=t=2$, and $\mathbb{I}$ denotes the $2\times 2$ identity matrix. In other
words, recalling (\ref{norm-J2}), one can conclude that the squared norm $%
\left\vert x\right\vert ^{2}$ of $x$ (as a vector in $\mathbf{M}^{2,2}$) is
given by the quadratic norm of $x$ as an element (\ref{X-call}) of $J_{2}^{%
\mathbb{C}_{s}}$ itself :%
\begin{equation}
\left\vert x\right\vert ^{2}=x^{a}x^{b}\eta _{ab}=\text{det}\mathcal{X}=%
\mathbf{N}\left( x\right) .  \label{vector-norm}
\end{equation}

Let us now consider the following transformations :

\begin{equation}
\begin{array}{rcl}
J_{2}^{\mathbb{C}_{s}} & \rightarrow  & J_{2}^{\mathbb{C}_{s}}, \\[3mm]
\mathcal{X} & \mapsto  & \lambda ^{\dag }\mathcal{X}\lambda =:\mathcal{X}%
^{\prime }\,,\,\,\,\,\,\lambda \in M_{2}(\mathbb{\mathbb{C}}_{s}),%
\end{array}
\label{transfo-2}
\end{equation}%
where $\dag $ stands for transposition times conjugation (\ref{conjug-Cs})
in the underlying split algebra $\mathbb{\mathbb{C}}_{s}$. \textit{Klein
transformations} are defined as those transformations (\ref{transfo-2}) in
which $\lambda \in \mathrm{SL}(2,\mathbb{C}_{s})$; it is then immediate to
realize that such transformations induce orthogonal transformations in $%
\mathbf{M}^{2,2}$, since they do preserve the determinant of $\mathcal{X}$,
and thus $\left\vert x\right\vert ^{2}$. In particular, $\mathrm{SL}(2,%
\mathbb{C}_{s})\cong \mathrm{SL}(2,\mathbb{R})\times \mathrm{SL}(2,\mathbb{R}%
)$ doubly covers $\mathrm{SO}(2,2)$, and it is then possible to identify it
(or, more precisely, its identity-connected component) with the \textit{Spin
}group $\mathrm{Spin}(2,2)$, which we anticipated above to be named \textit{%
Klein group} in $4$ dimensions. In other words, $\mathrm{SL}(2,\mathbb{C}%
_{s})$ acts naturally on $J_{2}^{\mathbb{C}_{s}}$ as the spin covering of $%
\mathrm{SO}(2,2)$. Thus, the following group isomorphisms hold:%
\begin{equation}
\mathrm{Spin}(2,2)\cong \mathrm{SL}(2,\mathbb{C}_{s})\cong \mathrm{SL}(2,%
\mathbb{R})\times \mathrm{SL}(2,\mathbb{R}).  \label{isso}
\end{equation}%
\medskip

As we have mentioned above, in signature $\left( 2,2\right) $ spinors are
Majorana, and they are identified with vectors in $\mathbb{C}_{s}^{2}$. We
identify them with the vector representation of $\mathrm{SL}(2,\mathbb{C}%
_{s})$, \emph{i.e.} $\mathbb{C}_{s}^{2}$. It is here instructive to observe
that, as an $\mathrm{SL}(2,\mathbb{C}_{s})$-module, $\mathbb{C}_{s}^{2}$ is
\textsl{not} irreducible. This can be realized by decomposing every vector
in $\mathbb{C}_{s}^{2}$ according to (\ref{decomposition}) as
\begin{equation}
\underbrace{\left(
\begin{matrix}
a \\
c%
\end{matrix}%
\right) }_{\psi }=\underbrace{\left(
\begin{matrix}
\alpha _{+} \\
\gamma _{+}%
\end{matrix}%
\right) }_{\psi _{\mathcal{E}}}\mathcal{E}+\underbrace{\left(
\begin{matrix}
\alpha _{-} \\
\gamma _{-}%
\end{matrix}%
\right) }_{\psi _{\overline{\mathcal{E}}}}\overline{\mathcal{E}}%
\,,\,\,\,\,\,\,\,\,\,\alpha _{\pm },\gamma _{\pm }\in \mathbb{R};
\end{equation}%
analogously, any element of $\mathrm{M}_{2}(\mathbb{C}_{s})$ can be split
as follows :
\begin{equation}
\underbrace{\left(
\begin{matrix}
a & b \\
c & d%
\end{matrix}%
\right) }_{M}=\underbrace{\left(
\begin{matrix}
\alpha _{+} & \beta _{+} \\
\gamma _{+} & \delta _{+}%
\end{matrix}%
\right) }_{M_{\mathcal{E}}}\mathcal{E}+\underbrace{\left(
\begin{matrix}
\alpha _{-} & \beta _{-} \\
\gamma _{-} & \delta _{-}%
\end{matrix}%
\right) }_{M_{\overline{\mathcal{E}}}}\overline{\mathcal{E}}.
\end{equation}

Consider now a matrix $M=M_{\mathcal{E}}\mathcal{E}+M_{\overline{\mathcal{E}}%
}\overline{\mathcal{E}}\in \mathrm{SL}(2,\mathbb{C}_{s})$; then, $2M_{%
\mathcal{E}}\in \mathrm{SL}(2,\mathbb{R})$ and $2M_{\overline{\mathcal{E}}%
}\in \mathrm{SL}(2,\mathbb{R})$ and every $\mathrm{SL}(2,\mathbb{C}_{s})$%
-module $\psi \in \mathbb{C}_{s}^{2}$ splits into two irreducible submodules
on which $M$ acts by an $\mathrm{SL}(2,\mathbb{R})$ matrix. 
%
To see this, we observe that $\det M=2\det M_{\mathcal{E}}\,\mathcal{E}%
+2\det M_{\overline{\mathcal{E}}}\,\overline{\mathcal{E}}$, from which one
obtains that the unitarity of $M$ implies $\det M_{\mathcal{E}}=\det M_{%
\overline{\mathcal{E}}}=\frac{1}{4}$, and thus $2M_{\mathcal{E}}$ and $2M_{%
\overline{\mathcal{E}}}$ are $\mathrm{SL}(2,\mathbb{R})$-matrices. Then, the
action on any spinors splits as
\begin{equation}
M\psi =(2M_{\mathcal{E}})\,\psi _{\mathcal{E}}+(2M_{\overline{\mathcal{E}}%
})\,\psi _{\overline{\mathcal{E}}}\,.
\end{equation}%
Consistent with (\ref{Weyl}), we thus identify $\psi _{\mathcal{E}}$ and $%
\psi _{\overline{\mathcal{E}}}$ with the Majorana-Weyl spinors $\psi ^{+}$
resp. $\psi ^{-}$ of opposite chirality.

\subsection{\label{reps}Spinor Orbits and Representatives}

Let us now discuss how the linear action of $\mathrm{Spin}(2,2)\left( \times
\mathrm{SO}(1,1)\right) $ on the spinor $\psi =(\mathbf{2},\mathbf{1}%
)_{+}\oplus (\mathbf{1},\mathbf{2})_{-}$ (or, equivalently, on its conjugate
$\overline{\psi }=\left( \mathbf{2,1}\right) _{-}+\left( \mathbf{1},\mathbf{2%
}\right) _{+}$) determines the \textsl{stratification} of the corresponding
spinor representation space $S$ into orbits. The crucial outcome of our
analysis (in agreement with literature; \textit{cfr. e.g.} \cite%
{Bryant-1,Bryant-2}, and Refs. therein) is that $\mathrm{Spin}(2,2)\cong
\mathrm{SL}(2,\mathbb{C}_{s})$ (\textit{cfr.} (\ref{isso})) does \textsl{not}
act transitively on $\mathbb{C}_{s}^{2}$.

We start by noting that the orbit of\footnote{%
The upperscript \textquotedblleft $t$" denotes transposition.} $%
e_{1}:=(1,0)^{t}\in \mathbb{C}_{s}^{2}$ contains all elements of the form $%
(a,c)^{t}$ with $a$ \textsl{and/or} $c$ \textsl{invertible}; in fact :
\begin{eqnarray}
\begin{pmatrix}
a & 0 \\
c & a^{-1}%
\end{pmatrix}%
\begin{pmatrix}
1 \\
0%
\end{pmatrix}
&=&%
\begin{pmatrix}
a \\
c%
\end{pmatrix}%
, \\
\begin{pmatrix}
a & -c^{-1} \\
c & 0%
\end{pmatrix}%
\begin{pmatrix}
1 \\
0%
\end{pmatrix}
&=&%
\begin{pmatrix}
a \\
c%
\end{pmatrix}%
.
\end{eqnarray}%
Thus, we are henceforth going to deal only with elements $(a,c)^{t}\in
\mathbb{C}_{s}^{2}$ with \textsl{both} $a$ and $c$ \textsl{non-invertible}.
By recalling the remark below (\ref{Cs-inv}) and Eq. (\ref{res}), this
amounts to consider \textsl{both} $a$ and $c$ either zero or $u\mathcal{E}$
or $u^{\prime }\overline{\mathcal{E}}$, with $u,u^{\prime }\in \mathbb{C}%
_{s}^{\times }$ and $\mathcal{E}:=1+j$.

We notice that
\begin{equation}
\begin{pmatrix}
u^{-1} & 0 \\
0 & u%
\end{pmatrix}%
\begin{pmatrix}
u\mathcal{E} \\
u^{\prime }\mathcal{E}%
\end{pmatrix}%
=%
\begin{pmatrix}
\mathcal{E} \\
uu^{\prime }\mathcal{E}%
\end{pmatrix}%
;
\end{equation}%
furthermore, $(\mathcal{E},v\mathcal{E})^{t}$ lies in the orbit of $(%
\mathcal{E},\mathcal{E})^{t}$, because
\begin{equation}
\begin{pmatrix}
1 & 0 \\
1-v & 1%
\end{pmatrix}%
\begin{pmatrix}
\mathcal{E} \\
v\mathcal{E}%
\end{pmatrix}%
=%
\begin{pmatrix}
\mathcal{E} \\
\mathcal{E}%
\end{pmatrix}%
.
\end{equation}

Therefore, up to conjugation in $S\cong \mathbb{C}_{s}^{2}$ (induced by the
conjugation in $\mathbb{C}_{s}$, and mapping the spinor $\psi $ into its
conjugate $\overline{\psi }$), in $S$ one needs to consider (besides $%
(0,0)^{t}$ and $(1,0)^{t}$) only the following elements:
\begin{equation}
\mathbf{1}:%
\begin{pmatrix}
0 \\
\mathcal{E}%
\end{pmatrix}%
,\quad \mathbf{2}:%
\begin{pmatrix}
\mathcal{E} \\
0%
\end{pmatrix}%
,\quad \mathbf{3}:%
\begin{pmatrix}
\mathcal{E} \\
\overline{\mathcal{E}}%
\end{pmatrix}%
,\quad \mathbf{4}:%
\begin{pmatrix}
\mathcal{E} \\
\mathcal{E}%
\end{pmatrix}%
;\quad   \label{I-IV}
\end{equation}%
in other words, one can disregard the multiplication by invertible split
complex numbers (as well as the conjugation in $\mathbb{C}_{s}^{2}$) when
dealing with the stratification of the spinor representation space $S$.

By definition of group orbit, in order to establish the \textit{%
stratification structure} of $\mathbb{C}_{s}^{2}$ under the action of the
Klein group in four dimensions, we have to determine which elements in $%
\mathbb{C}_{s}^{2}$ are connected through the action of an element $g\in
\mathrm{SL}(2,\mathbb{C}_{s})$. Let us then analyze the elements listed in (%
\ref{I-IV}) :

\begin{enumerate}
\item This element belongs to the orbit of $(\mathcal{E},\mathcal{E})^{t}$,
because :
\begin{equation}
\begin{pmatrix}
(1+\overline{\mathcal{E}})/2 & (1-\mathcal{E})/2 \\
(1-\overline{\mathcal{E}})/2 & (1+\mathcal{E})/2%
\end{pmatrix}%
\begin{pmatrix}
\mathcal{E} \\
\mathcal{E}%
\end{pmatrix}%
=%
\begin{pmatrix}
0 \\
2\mathcal{E}%
\end{pmatrix}%
.  \label{ep}
\end{equation}

\item A similar argument also shows that $(\mathcal{E},0)^{t}$ is in the
orbit of $(\mathcal{E},\mathcal{E})^{t}$.

\item Quite surprisingly, the element $(\mathcal{E},\overline{\mathcal{E}}%
)^{t}$ can be proved to lie in the orbit of $(1,0)^{t}$, because :
\begin{equation}
\begin{pmatrix}
\mathcal{E} & -1/2 \\
\overline{\mathcal{E}} & 1/2%
\end{pmatrix}%
\begin{pmatrix}
1 \\
0%
\end{pmatrix}%
=%
\begin{pmatrix}
\mathcal{E} \\
\overline{\mathcal{E}}%
\end{pmatrix}%
.
\end{equation}

\item There exists \textsl{no} transformation of $SL(2,\mathbb{C}_{s})$
connecting $(\mathcal{E},\mathcal{E})^{t}$ to $(1,0)^{t}$. In fact, if this
were the case, one would have
\begin{equation}
\begin{pmatrix}
a & b \\
c & d%
\end{pmatrix}%
\begin{pmatrix}
1 \\
0%
\end{pmatrix}%
=%
\begin{pmatrix}
\mathcal{E} \\
\mathcal{E}%
\end{pmatrix}%
,
\end{equation}%
hence $a=c=\mathcal{E}$. This \textsl{cannot} be, since otherwise the
determinant $ad-bc=\mathcal{E}(d-b)$ would be a zero divisor $\blacksquare $
\end{enumerate}

From this analysis, it follows that the orbits of $\mathbb{C}_{s}^{2}$ under
the action of the Klein group $\mathrm{SL}(2,\mathbb{C}_{s})$ (up to
conjugation in $\mathbb{C}_{s}^{2}$, equivalent to conjugation in $S$) are
characterized by one of the following three well-defined \textit{%
representatives} :
\begin{equation}
\begin{pmatrix}
0 \\
0%
\end{pmatrix}%
,\quad
\begin{pmatrix}
1 \\
0%
\end{pmatrix}%
,\quad
\begin{pmatrix}
\mathcal{E} \\
\mathcal{E}%
\end{pmatrix}%
.
\end{equation}

Thus, besides the trivial orbit (given by the origin $\left( 0,0\right) ^{t}$
of $\mathbb{C}_{s}^{2}$), $(1,0)^{t}$ and $\left( \mathcal{E},\mathcal{E}%
\right) ^{t}$ (or equivalently $\left( \mathcal{E},0\right) ^{t}$) are
well-defined representatives of the orbit stratification. In particular, the
representative $(1,0)^{t}$ is stabilized by any matrix of $\mathrm{SL}(2,%
\mathbb{C}_{s})$ of the form
\begin{equation}
\left(
\begin{matrix}
1 & b \\
0 & 1%
\end{matrix}%
\right) \cong \mathbb{C}_{s}\cong \mathbb{R}^{2},
\end{equation}%
while the stabilizer of $(\mathcal{E},0)^{t}$ reads
\begin{equation}
M_{\overline{\mathcal{E}}}\overline{\mathcal{E}}+\left(
\begin{matrix}
\frac{1}{2} & \beta _{+} \\
0 & \frac{1}{2}%
\end{matrix}%
\right) \mathcal{E}\cong \mathrm{SL}(2,\mathbb{R})\ltimes \mathbb{R}.
\end{equation}

Summarizing, we obtained

\begin{eqnarray}
\mathcal{O}_{(1,0)^{t}} &\cong &\frac{\mathrm{Spin}(2,2)}{\mathbb{R}^{2}},~%
\text{dim}_{\mathbb{R}}=4;  \label{ps-2-2} \\
\mathcal{O}_{(\mathcal{E},0)^{t}} &\cong &\frac{\mathrm{Spin}(2,2)}{\mathrm{%
SL}(2,\mathbb{R})\ltimes \mathbb{R}},~\text{dim}_{\mathbb{R}}=2.
\label{ps-3-2}
\end{eqnarray}%
Since dim$_{\mathbb{R}}\mathcal{O}_{(1,0)^{t}}=$dim$_{\mathbb{R}}S=4$, such
an orbit can be regarded as the \textsl{generic} one; consequently, $%
\mathcal{O}_{(\mathcal{E},0)^{t}}$ is the \textsl{non-generic} spinor orbit.

\section{Klein-Conformal group and $\mathrm{Spin}(3,3)$}

We are now going to consider the \textit{Klein-ambient space} $\mathbf{M}^{3,3}\cong
\mathbb{R}^{3,3}$ and the corresponding \textit{Klein group} in $6$
dimensions, namely $\mathrm{Spin}(3,3)$. In this case, $s=t=3$, and thus
again $\rho =0$. In turn, $\mathrm{Spin}(3,3)$ can also be regarded as the
\textit{conformal}\footnote{%
This observation will also give rise to the chain of isomorphisms (\ref%
{iso-3}) (holding both at Lie algebra and at Lie group level). It can be
traced back to the symmetry of the \textit{doubly-split} Magic Square of
order $2$ \cite{BS-1,BS-2,MS-2-Groups}.} group of $\mathbf{M}^{2,2}\cong
\mathbb{R}^{2,2}$ itself.

In complete analogy with the treatment of $\mathrm{Spin}(2,2)$ given above,
one can identify a vector $x^{A}=(x^{1},\cdots ,x^{6})$ ($A=1,...,6$) in $%
\mathbf{M}^{3,3}$ with an element of the quadratic simple Jordan algebra $%
J_{2}^{\mathbb{H}_{s}}$over split quaternions $\mathbb{H}_{s}$, by
rearranging the vector components as entries of the $2\times 2$ Hermitian
matrix
\begin{equation}
\mathcal{V}=%
\begin{pmatrix}
\hat{x}_{+} & z^{\ast } \\
z & \hat{x}_{-}%
\end{pmatrix}%
\in J_{2}^{\mathbb{H}_{s}},  \label{V-J2-Hs}
\end{equation}%
where $z:=x^{5}+jx^{1}+kx^{4}+(kj)x^{2}\in \mathbb{H}_{s}$, $\mathbb{R}\ni
\hat{x}_{\pm }:=x^{3}\pm x^{6}$ , and the star denoting the conjugation in $%
\mathbb{H}_{s}$ (\textit{cfr.} (\ref{conjug-Hs})). By recalling the
definition (\ref{norm-Hs}), the quadratic form associated to the metric $%
\eta _{AB}=\eta ^{AB}$ of signature $(3,3)$ (given by (\ref{metric}) with $%
s=t=3$) of $\mathbf{M}^{3,3}$ is then obtained by computing\footnote{%
It should be here remarked that the determinant of $2\times 2$ Hermitian
matrices with $\mathbb{H}$- or $\mathbb{H}_{s}$- valued entries is well
defined \cite{McCrimmon}.}
\begin{equation}
\text{det}\mathcal{V}=\left( x^{3}\right) ^{2}-\left( x^{6}\right)
^{2}-|z|^{2}=\left( x^{1}\right) ^{2}+\left( x^{2}\right) ^{2}+\left(
x^{3}\right) ^{2}-\left( x^{4}\right) ^{2}-\left( x^{5}\right) ^{2}-\left(
x^{6}\right) ^{2}:=\eta _{AB}x^{A}x^{B},
\end{equation}%
or equivalently
\begin{equation}
\mathcal{V}\widetilde{\mathcal{V}}=\widetilde{\mathcal{V}}\mathcal{V}=-\eta
_{AB}x^{A}x^{B}\mathbb{I}\,,
\end{equation}%
where $\widetilde{\mathcal{V}}$ is the \textit{trace reversal} of $\mathcal{V%
}$. In other words, recalling (\ref{norm-J2}), one can conclude that the
squared norm $\left\vert x\right\vert ^{2}$ of $x$ (as a vector in $\mathbf{M%
}^{3,3}$) is given by the quadratic norm of $x$ as an element (\ref{V-J2-Hs}%
) of $J_{2}^{\mathbb{H}_{s}}$ itself :%
\begin{equation}
\left\vert x\right\vert ^{2}=x^{A}x^{A}\eta _{AB}=\text{det}\mathcal{V}=%
\mathbf{N}\left( x\right) .
\end{equation}

Let us now consider the following transformations :
\begin{equation}
\begin{array}{rcl}
J_{2}^{\mathbb{H}_{s}} & \rightarrow & J_{2}^{\mathbb{H}_{s}}, \\[3mm]
\mathcal{V} & \mapsto & \lambda ^{\dag }\mathcal{V}\lambda =:\mathcal{V}%
^{\prime }\,,\,\,\,\,\,\lambda \in M_{2}(\mathbb{H}_{s}),%
\end{array}
\label{transfo-3}
\end{equation}%
where $\dag $ stands for transposition times conjugation (\ref{conjug-Hs})
in the underlying split algebra $\mathbb{\mathbb{H}}_{s}$. \textit{%
Klein-conformal transformations} are defined as those transformations (\ref%
{transfo-3}) in which $\lambda \in \mathrm{SL}(2,\mathbb{H}_{s})$, where the
special linear group is defined as (recall (\ref{transfo}))
\begin{equation}
\mathrm{SL}(2,\mathbb{H}_{s}):=\{M\in M_{2}({\mathbb{H}_{s}})\,|\,\text{det}%
(Z(M))=1\},
\end{equation}%
or equivalently (recall (\ref{epsilon}))
\begin{equation}
\mathrm{SL}(2,\mathbb{H}_{s}):=\left\{ M\in \mathrm{SL}(4,{\mathbb{C}_{s}}%
)\,\left\vert\, \overline{M}%
\begin{pmatrix}
\epsilon & \mathbf{0} \\
\mathbf{0} & \epsilon%
\end{pmatrix}%
=%
\begin{pmatrix}
\epsilon & \mathbf{0} \\
\mathbf{0} & \epsilon%
\end{pmatrix}%
M\right. \right\} .
\end{equation}%
It is then immediate to realize that such transformations induce orthogonal
transformations in $\mathbf{M}^{3,3}$ (and correspondingly \textsl{conformal}
transformations in $\mathbf{M}^{2,2}$), since they do preserve det$\mathcal{V%
}$ and thus $\left\vert x\right\vert ^{2}$. In particular, $\mathrm{SL}(2,%
\mathbb{H}_{s})$ doubly covers $\mathrm{SO}(3,3)$, and it is then possible
to identify it (or, more precisely, its identity-connected component) with
the \textit{Spin }group $\mathrm{Spin}(3,3)$. In other words, $\mathrm{SL}(2,%
\mathbb{H}_{s})$ acts naturally on $J_{2}^{\mathbb{H}_{s}}$ as the spin
covering of $\mathrm{SO}(3,3)$. This establishes the group isomorphism%
\begin{equation}
\mathrm{Spin}(3,3)\cong \mathrm{SL}(2,\mathbb{H}_{s}).\newline
\label{iso-2}
\end{equation}

\subsection{A Further Group Isomorphism}

For the subsequent treatment, we find convenient to present also another
isomorphism involving $\mathrm{Spin}(3,3)$, namely\footnote{%
The tilde in $\widetilde{\mathrm{Sp}}(4,\mathbb{C}_{s})$ denotes the
peculiar definition (\ref{def-Sp4}) - after \cite{BS-1,BS-2} - of the
symplectic group by the matrix Hermitian-conjugate (and not by the matrix
transpose, as usually done).}%
\begin{equation}
\mathrm{Spin}(3,3)\cong \widetilde{\mathrm{Sp}}(4,\mathbb{C}_{s}).
\label{iso-1}
\end{equation}%
In order to prove it, we start from the $4\times 4$ matrix given by (\ref%
{app-1}) in the App. \ref{App-Sympl}, which we report below for
convenience's sake : \texttt{\ }
\begin{equation}
\mathbb{X}:=%
\begin{pmatrix}
\hat{x}_{+}\epsilon  & \mathcal{X}\epsilon  \\[3mm]
-\widetilde{\mathcal{X}}\epsilon  & -\hat{x}_{-}\epsilon
\end{pmatrix}%
.
\end{equation}%
Then, one can compute that
\begin{equation}
\mathbb{X}^{\dag }\Omega \mathbb{X}=\eta _{AB}x^{A}x^{B}\Omega ,
\end{equation}%
where $\dag $ stands for transposition times conjugation (\ref{conjug-Cs})
in $\mathbb{\mathbb{C}}_{s}$, and $\Omega $ here denotes for the $4\times 4$
symplectic metric (recall (\ref{epsilon}))%
\begin{equation}
\Omega =%
\begin{pmatrix}
0 & \mathbb{I} \\
-\mathbb{I} & 0%
\end{pmatrix}%
=\mathbb{I}\otimes \epsilon .  \label{Omega-4x4}
\end{equation}%
Therefore, one can define the symplectic group \textsl{\`{a} la Barton and
Sudbery} \cite{BS-1,BS-2}:
\begin{equation}
\widetilde{\mathrm{Sp}}(4,\mathbb{C}_{s}):=\{M\in \mathrm{SL}(4,\mathbb{C}%
_{s})\,|\,M^{\dag }\Omega M=\Omega \},  \label{def-Sp4}
\end{equation}%
and any transformation of the form
\begin{equation}
\mathbb{X}\rightarrow \mathbb{X}=\lambda \mathbb{X}\lambda ^{t},\,\,\,\text{%
with}\,\,\,\,\lambda \in \widetilde{\mathrm{Sp}}(4,\mathbb{C}_{s}),
\end{equation}%
preserves the (squared) norm $\left\vert x\right\vert ^{2}$ in $\mathbf{M}%
^{3,3}$ and induces a \textsl{Klein-conformal }transformation in $\mathbf{M}%
^{2,2}$, thus providing an alternative realization of $\mathrm{Spin}(3,3)$,
and then inducing the isomorphism (\ref{iso-1}) $\blacksquare $

We thus obtain the following chain of group isomorphisms:%
\begin{equation}
\mathrm{Spin}\left( 3,3\right)\cong \mathrm{%
SL}(2,\mathbb{H}_{s})\cong \widetilde{\mathrm{Sp}}(4,\mathbb{C}_{s}).
\label{iso-3}
\end{equation}
We have already discussed in Section \ref{spinors} that spinors of $\mathrm{Spin}\left( 3,3\right)$ can be interpreted as vectors of $\mathbb{H}_{s}^2\cong \mathbb{C}_{s}^4$ on which, in analogy with the $\mathrm{Spin}\left( 2,2\right)$ case, $\mathrm{%
SL}(2,\mathbb{H}_{s})$ does not act transitively; this determines the stratification into orbits. Even if the stratification reveals to be more evident using the special linear group over split quaternions, in the next section we will use instead the symplectic group over split complexes to realize the Klein superspace
as the space of $2|0$ totally
isotropic subspaces in $\mathbb{C}_{s}^{4|1}$, \emph{i.e.} the Lagrangian superspace.

We will focus on the case in which the representative super plane is given by a pair of generic vectors of $\mathbb{C}_{s}^{4|1}$ whose even part is a generic spinor. One has of course the possibility to choose other isotropic subspaces as representative given by other combinations (namely, non generic-generic or non generic-non generic) of spinors. Since the action of the Klein group stratifies the spinor space, we expect to obtain different and intriguing constructions. We leave a detailed analysis for a future project, while in this paper we focus on the generic-generic case for spinor representatives.

\section{Klein  and Klein-Conformal $\mathcal{N}=1$
Superspaces\label{supermink}}

We can now proceed to construct the $\mathcal{N}=1$ Klein-conformal and
Klein superspaces in $D=(2,2)$. Supermanifolds, and in particular the $%
\mathcal{N}=1$ Minkowski and conformal superspaces in $D=(3,1)$, have been
studied intensively in the past years. A thorough account of such a broad
field of investigation lies well beyond the scope of this paper; we here
confine ourselves to addressing the interested reader to \cite{flmink}, and
Refs. therein, for an exhaustive bibliography.

In order to construct the $\mathcal{N}=1$ Klein-conformal and Klein
superspaces in $D=(2,2)$, we will exploit a procedure which is very similar
to the one of \cite{FL-1}; however, some extra attention should be paid in
the definition of the \textsl{functor of points} of a $\mathbb{C}_{s}$%
-group. We could give our definitions in full generality, but for clarity's
sake we do prefer to adapt them to our specific framework.
We have provided App. \ref{sgeo-app} for the basic facts
of supergeometry and supergroups; for more details
on the technicalities involved, we
address the reader \textit{e.g.} to Ch.
10 of \cite{ccf}.

The $A$-points of the general linear supergroup over $\C_s$ are given by
(see App. \ref{sgeo-app} (\ref{glfun})):

\begin{equation}
\mathrm{GL}(m|n)(A) 
=\left\{ \begin{pmatrix}
a & \alpha  \\
\beta  & b%
\end{pmatrix}
\right\}\,=\, \mathrm{Hom}_{\mathrm{(salg)}}(
\C_s[\mathrm{GL}(m|n)],A),
\end{equation}%
where
$a$, $b$, $\alpha $, $\beta $ are matrices with entries in $A$ (roman
and greek lowercase letters denote even resp. odd entries throughout), and $a
$ and $b$ are invertible.

If we regard $\rGL(m|n)$ as a real supergroup, we can define its $A$-points
(where here $A$ is a real superalgebra):
\begin{equation}
(\rGL(m|n)(A))_{\mathbb{R}}=\Hom_{\salg}(\C_{s}[[x_{ij},\xi _{kl}][\det
(x_{ij})_{1\leq i,j\leq m}^{-1},\det (x_{ij})_{m+1\leq i,j\leq
m+n}^{-1}],A\otimes \C_{s})
\end{equation}
(see App. \ref{sgeo-app} (\ref{cs-real})).

We now define, in complete analogy to \cite{FL-1}, the \textsl{%
symplectic orthogonal supergroup}\textit{\ }$\widetilde{\mathrm{SpO}}(4|1)$
as the (real) subsupergroup of $\mathrm{GL}(m|n)_{%
\mathbb{R}}$ given as:

\begin{equation*}
\widetilde{\mathrm{SpO}}{(4|1)}(A)\,=\,\{\Lambda \in (\mathrm{GL}{(4|1)}
)_{\mathbb{R}}(A)\,|\,\Lambda ^{\dag }\mathbb{J}\Lambda =%
\mathbb{J}\},\qquad \text{with}\qquad \mathbb{J}=%
\begin{pmatrix}
0 & \mathbb{I} & 0 \\
-\mathbb{I} & 0 & 0 \\
0 & 0 & 1%
\end{pmatrix}%
=%
\begin{pmatrix}
\Omega & 0 \\
0 & 1%
\end{pmatrix}%
\,.
\end{equation*}%
where $\Lambda ^{\dagger }:=\overline{\Lambda }^{t}$ (with $t$ here denoting
the supertranspose) and the conjugation is consistently understood in $%
\mathbb{C}_{s}$, as detailed in the treatment above. If

\begin{equation}
\Lambda =%
\begin{pmatrix}
B & \alpha \\
\beta & u%
\end{pmatrix}%
,\quad B=%
\begin{pmatrix}
a & b \\
c & d%
\end{pmatrix}%
,\quad \beta =(\beta _{1},\beta _{2}),\quad \alpha =(\alpha _{1},\alpha
_{2})^{t},
\end{equation}

with $\beta _{i}$, $\alpha _{i}\in A^{2}$ ($i=1,2$), from the
condition%
\begin{equation}
\Lambda ^{\dag }\mathbb{J}\Lambda =\mathbb{J}
\end{equation}%
one obtains the following set of equations :

\begin{equation}
\left\{
\begin{array}{c}
B^{\dagger }JB+\beta ^{\dagger }\beta =J; \\
\\
B^{\dagger }J\alpha +\beta ^{\dagger }u=0; \\
\\
-\alpha ^{\dagger }JB+u^{\dagger }\beta =0; \\
\\
-\alpha ^{\dagger }J\alpha +u^{\dagger }u=1.%
\end{array}%
\right. \,\iff \,\left\{
\begin{array}{c}
a^{\dagger }c-c^{\dag }a+\beta _{1}^{\dagger }\beta _{1}=0; \\
a^{\dagger }d-c^{\dagger }b+\beta _{1}^{\dagger }\beta _{2}=\mathbb{I}; \\
b^{\dagger }c-d^{\dagger }a+\beta _{2}^{\dagger }\beta _{1}=-\mathbb{I}; \\
b^{\dagger }d-d^{\dagger }b+\beta _{2}^{\dagger }\beta _{2}=0; \\
-c^{\dagger }\alpha _{1}+a^{\dagger }\alpha _{2}+\beta _{1}^{\dagger }u=0;
\\
-d^{\dagger }\alpha _{1}+b^{\dagger }\alpha _{2}+\beta _{2}^{\dagger }u=0;
\\
\alpha _{2}^{\dagger }\alpha _{1}-\alpha _{1}^{\dagger }\alpha
_{2}+u^{\dagger }u=1.%
\end{array}%
\right.   \label{spo-eq}
\end{equation}%
\medskip

We now consider the (real) supermanifold $\mathcal{L}$ of $2|0$ totally
isotropic subspaces in $\mathbb{C}_{s}^{4|1}$. Let
us take $\{\mathbf{e}_{1},\mathbf{e%
}_{2},\mathbf{e}_{3},\mathbf{e}_{4},\epsilon\}$ the canonical
basis for $\mathbb{C}_{s}^{4|1}$.
We define $\mathcal{L}$ as the orbit of the super subspace
$\mathrm{span}_{\C_s} \{{\mathbf{e}}_{1},{\mathbf{e}}_{2}\}$
under the natural action of the real supergroup
$\widetilde{\mathrm{SpO}}{(4|1)}$.
This is a supermanifold, and if $A$ is a
local $\mathbb{C}_{s}$-superalgebra, one obtains

\begin{equation}
\mathcal{L}(A)
=\left\{
\begin{pmatrix}
a \\
c \\
\beta _{1}%
\end{pmatrix}%
\,\Big|\,a^{\dagger }c-c^{\dagger }a+\beta _{1}^{\dagger }\beta
_{1}=0\right\} /\mathrm{GL}_{2}(A)\,.  \label{slag-eq}
\end{equation}%
It should be here
stressed that $A$ needs to be taken \textsl{local} in order to express in an
easier way the action of $\widetilde{\mathrm{SpO}}{(4|1)}$
on $\mathcal{L}$; we address the
reader to Chs. 2 and 4 of \cite{flmink} for a detailed treatment of this
technical point.
\smallskip

\textbf{Remark}.  The real supergroup $\widetilde{\mathrm{SpO}}{(4|1)}$
does \textsl{not} act transitively on the superspace $\mathbb{C}_{s}^{4|1}$;
in the standard (\textit{i.e.}, non-super) case, we have mentioned such a feature in the previous section and in Sec. \ref{reps} for the Klein case. However, this fact will not
influence our treatment, since we realize the Klein $\mathcal{N}=1$
superspace as an \textsl{open} inside the
$\widetilde{\mathrm{SpO}}{(4|1)}$-orbit $\mathcal{L}$ of
$\mathrm{span}_{\C_s} \{{\mathbf{e}}_{1},{\mathbf{e}}_{2}\}$,\emph{i.e.} of the generic-generic spinor case.
\smallskip

We consider the open subset of $\mathcal{L}$ consisting of those subspaces
corresponding to $a$ invertible. We call it $\mathbf{M}^{2,2|1}$ : it will
be our model for the $D=(2,2)$ $\mathcal{N}=1$ \textsl{Klein superspace},
while $\mathcal{L}$ is topologically the compactification of $\mathbf{M}%
^{2,2|1}$, and it is the $D=(2,2)$ $\mathcal{N}=1$ \textsl{Klein-conformal
superspace}. By multiplying by a suitable element of $\mathrm{GL}_{2}(A)$ we
have:

\begin{equation}
\mathbf{M}^{2,2|1}(A)=\left\{
\begin{pmatrix}
\mathbb{I} \\
\mathcal{Y} \\
\zeta
\end{pmatrix}%
\,\left\vert \mathcal{Y}^{\dagger }=\mathcal{Y}+\zeta ^{\dagger }\zeta
\right. \right\} \,.\label{M2,2 1}
\end{equation}%
Here $A$ is a commutative superalgebra, not necessarily local as before.

Notice that $\mathcal{Y}=ca^{-1}$, $\zeta =\beta _{1}a^{-1}$ with respect to
the expression in (\ref{slag-eq}). Hence, the equation is obtained
immediately from (\ref{slag-eq}) by setting $a=1$. This is precisely the
condition found in \cite{flv} and in \cite{FL-1}. Furthermore, we remark that the relation:
\begin{equation*}
\mathcal{Y}^{\dagger }=\mathcal{Y}+\zeta ^{\dagger }\zeta
\end{equation*}%
for $\zeta =0$ reduces to the condition of $\mathcal{Y}$ to be Hermitian (in
the context of $\mathbb{C}_{s}$). A comparison with (\ref{X-call})
shows that this is precisely the condition for an element in $M_{2}(\mathbb{C%
}_{s})$ to belong to $\mathbf{M}^{2,2}$. Thus, the $\mathbb{C}_{s}$ points
of the supermanifold $\mathbf{M}^{2,2|1}$ coincide with the Klein space $%
\mathbf{M}^{2,2}$ discussed above, and this justifies the use of our
super-terminology.

We now proceed to examine the \textsl{Klein-Poincar\'{e} supergroup}, acting
on $\mathbf{M}^{2,2|1}$. We start by noticing that the supergroup functor

\begin{equation} 
\widehat{sKP}(A):=\left\{
\begin{pmatrix}
L & 0 & 0 \\
M & R & R\phi  \\
d\chi  & 0 & d%
\end{pmatrix}%
\right\} \subset \widetilde{\mathrm{SpO}}{(4|1)}(A) \label{sKP}
\end{equation}%
leaves $\mathbf{M}^{2,2|1}$ invariant
($A$ as usual is a commutative superalgebra).
This subgroup is representable (see App. \ref{sgeo-app} for
the definition of representable supergroup functor).
In fact 
the real superalgebra representing it is obtained as a quotient of
$\R[ \widetilde{\mathrm{SpO}}{(4|1)}]$, namely setting to zero those
generators corresponding to the positions where we have zeros for
the $A$-points in (\ref{sKP}).
Notice that its reduced group (see App.
\ref{sgeo-app}, (\ref{red-grp})) is the Klein-Poincar\'{e} group itself.

We then define $\widehat{sKP}$  as \textit{the Klein-Poincar%
\'{e} supergroup}.
Its $A$-points are given by (\ref{sKP}).
Applying the equations in (\ref{spo-eq}) to $\widehat{sKP}(A)$, one obtains

\begin{equation}
R=(L^{\dagger })^{-1},\quad \phi =\chi ^{\dagger },\quad
ML^{-1}=(ML^{-1})^{\dagger }+(L^{\dagger })^{-1}\chi ^{\dagger }\chi
L^{-1}\,,
\end{equation}%
yielding

\begin{equation}
\widehat{sKP}(A)=\left\{
\begin{pmatrix}
L & 0 & 0 \\
M & (L^{\dagger })^{-1} & (L^{\dagger })^{-1}\chi ^{\dagger } \\
d\chi  & 0 & d%
\end{pmatrix}%
\right\} .
\end{equation}%
\medskip Then, the action on $\mathbf{M}^{2,2|1}$ (\ref{M2,2 1}) can be
readily computed to yield :

\begin{equation}
\begin{array}{rclcl}
\widehat{sKP} & \times  & \mathbf{M}^{2,2|1} & \longrightarrow  & \mathbf{M}%
^{2,2|1} \\
&  &  &  &  \\
\scalebox{.8}{$\begin{pmatrix} L & 0 & 0 \\ M & (L^\dagger )^{-1} & (\chi
L^{-1})^\dag\  \\ d\chi & 0 & d \end{pmatrix}$} & , & \scalebox{.8}{$%
\begin{pmatrix} \mathbb{I}\\[2mm] \mathcal{Y} \\[2mm] \zeta \end{pmatrix} $}
& \mapsto  & \scalebox{.8}{$\begin{pmatrix} \mathbb{I}\\[2mm]
ML^{-1}+(L^\dag)^{-1}\mathcal{Y}L^{-1} +(\chi L^{-1})^\dag\zeta
L^{-1}\\[2mm] d\chi L^{-1}+d\zeta L^{-1} \end{pmatrix} $}.%
\end{array}%
\end{equation}%
\medskip
We end this Section with an important observation that relates our
construction of the Klein-Poincar\'{e} supergroup with our previous
treatment of spinors of $\mathrm{Spin}(2,2)$. The Klein-Poincar\'{e}
supergroup contains as its closed subgroup the \textsl{Klein supergroup},
whose functor of points is given by:
\begin{equation}
\widehat{sK}(A)=%
\begin{pmatrix}
(R^{\dagger })^{-1} & 0 & 0 \\
0 & R & R\phi  \\
d\phi ^{\dagger } & 0 & d%
\end{pmatrix}%
.
\end{equation}%
As for its counterpart in Lorentz signature, $\widehat{sK}$ is obtained from
$\widehat{sKP}$ by removing the inhomogeneous translational part given by $M$
(note we here use the variables $R=(L^{-1})^{\dagger }$, $\phi =\chi
^{\dagger }$). The corresponding Lie superalgebra reads
\begin{equation}
\mathrm{Lie}(\widehat{sK})=%
\begin{pmatrix}
-r^{\dagger } & 0 & 0 \\
0 & r & r\varphi  \\
\mathcal{D}\varphi ^{\dagger } & 0 & \mathcal{D}%
\end{pmatrix}%
.
\end{equation}%
If $\mathfrak{g}_{0}$ and $\mathfrak{g}_{1}$ respectively denote the even
and odd part of $\mathfrak{g}:=\mathrm{Lie}(\widehat{sK})$, there is a
natural action of $\mathfrak{g}_{0}$ on $\mathfrak{g}_{1}$. Indeed, in this
framework, it holds that $\mathfrak{g}_{0}=\mathfrak{g}_{0}^{\prime }\oplus
\mathbb{C}_{s}$, where $\mathfrak{g}_{0}^{\prime }$ is the Lie algebra of
the spin group $\mathrm{SL}_{2}(\mathbb{C}_{s})\cong \mathrm{Spin}(2,2)$ (%
\textit{cfr.} (\ref{isso})), and $\mathbb{C}_{s}$ corresponds to
dilatations. As one can readily check, the action of $\mathfrak{g}_{0}$ on
the odd part $r\varphi $ (that is $\mathfrak{g}_{1}$) is precisely the
spinor representation $\mathbb{C}_{s}^{2}$ studied in previous Sections.

\section*{Acknowledgments}

We would like to thank John C. Baez, Leron Borsten, Andrew Waldron,
Francesco Toppan and V. S. Varadarajan for useful discussions and
suggestions, and especially Tevian Dray for its kind help in understanding
the magic squares of Lie groups.

A.M. wishes to thank the Department of Mathematics at the University of
Bologna, for the kind hospitality during the realization of this work.

\appendix

\section{\label{App-Sympl}Symplectic Realization of $\mathrm{Spin}(3,3)$}

Consider the canonical basis $\{\mathbf{e}_{\mu }\}$ for $\mathbb{C}_{s}^{4}$, and $\{\mathbf{e%
}^{\mu }\}$ its dual basis ($\mu =1,...,4$). A natural inner product\ $%
<\bullet ,\bullet >$ in $\Lambda ^{2}\mathbb{C}_{s}^{4}$ can be defined as
follows
\begin{equation}
\begin{array}{rrrcl}
<\bullet ,\bullet > & : & \Lambda ^{2}\mathbb{C}_{s}^{4}\otimes \Lambda ^{2}%
\mathbb{C}_{s}^{4} & \rightarrow  & \mathbb{C}_{s}, \\
&  & x\wedge y\,,\,z\wedge w\,\,\, & \mapsto  & (x\wedge y\wedge z\wedge w)(%
\mathbf{e}^{1}\wedge \mathbf{e}^{2}\wedge \mathbf{e}^{3}\wedge \mathbf{e}%
^{4}).%
\end{array}%
\end{equation}%
Note that $\widetilde{\mathrm{Sp}}(4,\mathbb{C}_{s})$ (\ref{def-Sp4}) acts
in an obvious way on $\Lambda ^{2}\mathbb{C}_{s}^{4}$ preserving the inner
product $<\bullet ,\bullet >$.

We are now going to determine a \textsl{real} $6$-dimensional subspace of $%
\Lambda ^{2}\mathbb{C}_{s}^{4}$, which is stable under $\widetilde{\mathrm{Sp%
}}(4,\mathbb{C}_{s})$ and on which $<\bullet ,\bullet >$ takes real values.
To this aim, let us define the \textsl{symplectic} inner product
\begin{equation}
\begin{array}{rrrcl}
<\bullet ,\bullet >_{\Omega } & : & \mathbb{C}_{s}^{4}\otimes \mathbb{C}%
_{s}^{4} & \rightarrow & \mathbb{C}_{s}, \\
&  & x,y\,\,\, & \mapsto & y^{\dag }\Omega x,%
\end{array}%
\end{equation}%
where $\Omega $ is given by (\ref{Omega-4x4}).

Then, one can use $<\bullet ,\bullet >$ and $<\bullet ,\bullet >_{\Omega }$%
\thinspace\ in order to construct the isomorphisms $\phi :\Lambda ^{2}%
\mathbb{C}_{s}^{4}\xrightarrow{\sim}(\Lambda ^{2}\mathbb{C}_{s}^{4})^{\ast }$
and $\varphi :\mathbb{C}_{s}^{4}\xrightarrow{\sim}(\mathbb{C}_{s}^{4})^{\ast
}$. It is then possible to naturally identify $(\Lambda ^{2}\mathbb{C}%
_{s}^{4})^{\ast }\cong \Lambda ^{2}(\mathbb{C}_{s}^{\ast })^{4}$, and use it
to construct the $\widetilde{\mathrm{Sp}}(4,\mathbb{C}_{s})$-invariant
isomorphism of $\Lambda ^{2}\mathbb{C}_{s}^{4}$ into itself as $\Phi
:=\varphi ^{-1}\otimes \varphi ^{-1}\cdot \phi $.
This identifies a subspace of $\Lambda ^{2}\mathbb{C}_{s}^{4}$ on which $%
\Phi $ acts as the identity operator. A convenient basis of such a subspace
reads as follows :
\begin{equation}
\begin{array}{rclcrcl}
E_{1} & = & \frac{1}{\sqrt{2}}e_{1}\wedge e_{4}-\frac{1}{\sqrt{2}}%
e_{2}\wedge e_{3}; & \,\,\,\, & E_{4} & = & \frac{1}{\sqrt{2}}e_{1}\wedge
e_{4}+\frac{1}{\sqrt{2}}e_{2}\wedge e_{3}; \\
E_{2} & = & j\frac{1}{\sqrt{2}}e_{1}\wedge e_{3}+j\frac{1}{\sqrt{2}}%
e_{2}\wedge e_{4}; & \,\,\,\, & E_{5} & = & \frac{1}{\sqrt{2}}e_{2}\wedge
e_{4}-\frac{1}{\sqrt{2}}e_{1}\wedge e_{3}; \\
E_{3} & = & \frac{1}{\sqrt{2}}e_{1}\wedge e_{2}-\frac{1}{\sqrt{2}}%
e_{3}\wedge e_{4}; & \,\,\,\, & E_{6} & = & \frac{1}{\sqrt{2}}e_{1}\wedge
e_{2}+\frac{1}{\sqrt{2}}e_{3}\wedge e_{4}, \\
&  &  &  &  &  &
\end{array}%
\end{equation}%
and it can be checked that within the such a subspace the inner product $%
<\bullet ,\bullet >$ takes real values, and has signature $\left( 3,3\right)
$.

Therefore, any vector in this subspace can be represented as antisymmetric $%
4\times 4$ matrix of the form
\begin{eqnarray}
\mathbb{X} &:&=%
\begin{pmatrix}
0 & x_{3}+x_{6} & -x_{5}+jx_{2} & x_{1}+x_{4} \\
-x_{3}-x_{6} & 0 & x_{4}-x_{1} & x_{5}+jx_{2} \\
x_{5}-jx_{2} & -x_{4}+x_{1} & 0 & x_{6}-x_{3} \\
-x_{1}-x_{4} & -x_{5}-jx_{2} & x_{3}-x_{6} & 0%
\end{pmatrix}
\notag \\
&=&%
\begin{pmatrix}
\hat{x}_{+}\epsilon & \mathcal{X}\epsilon \\[3mm]
-\widetilde{\mathcal{X}}\epsilon & -\hat{x}_{-}\epsilon%
\end{pmatrix}%
=\epsilon \otimes
\begin{pmatrix}
\hat{x}_{+} & \mathcal{X} \\[3mm]
-\widetilde{\mathcal{X}} & -\hat{x}_{-}%
\end{pmatrix}%
,  \label{app-1}
\end{eqnarray}%
where in the last step definition (\ref{epsilon}) has been recalled, $\hat{x}%
_{\pm }:=x^{5}\pm x^{6}\in \mathbb{R}$, and $\mathcal{X}$, $\widetilde{%
\mathcal{X}}\in J_{2}^{\mathbb{C}_{s}}$ (\textit{cfr.} definitions (\ref%
{X-call})-(\ref{X-call-tilde})).

\section{Supergeometry} \label{sgeo-app}

In this appendix we recall few well known facts about superalgebras
and more in general supergeometry. We refer the reader to \cite{ccf} and the
references within for more details.

\medskip Let $k$ be a commutative algebra. For our purposes, it is enough to
consider the cases of $k=\R, \C,\C_s$.

\medskip A \textit{super vector space} is a $\Z/2\Z$-graded vector space $V
= V_0 \oplus V_1$; the elements of $V_0$ are called \textit{even} and
elements of $V_1$ are called \textit{odd}. Notice that a parity of a vector $%
v$, denoted by $p(v)$, is not defined in general, but, since any element may
be expressed as the sum of homogeneous ones, it suffices to consider only
homogeneous vectors in all of the statements relying on linearity.

The \textit{super dimension} of a super vector space $V$ is the pair $(p,q)$%
, where dim($V_{0}$)=$p$ and dim($V_{1}$)=$q$ as ordinary vector spaces.
When the dimension of $V$ is $p|q$, we can find a basis $\{e_{1},\ldots
,e_{p}\}$ of $V_{0}$ and a basis $\{\epsilon _{1},\ldots ,\epsilon _{q}\}$
of $V_{1}$ so that
\begin{equation}
V={\mathrm{span}}\{e_{1},\ldots ,e_{p},\epsilon _{1},\ldots ,\epsilon _{q}\}.
\end{equation}%
For us, the most relevant example is $\C_{s}^{4|1}={\mathrm{span}}%
\{e_{1},\ldots ,e_{4},\epsilon _{1}=:\epsilon\}$ (when there is just
one odd basis element we omit the numbering).

\medskip A \textit{superalgebra} over $k$ is a super vector space $A$
together with a multiplication preserving parity. $A$ is commutative if
\begin{equation}
xy=(-1)^{p(x)p(y)}yx
\end{equation}%
The prototype of a commutative superalgebra is the \textit{polynomial
superalgebra}, generated by the even indeterminates $t_{1},\dots ,t_{m}$,
which commute, and the odd ones $\theta _{1},\dots ,\theta _{n}$, which
anticommute: $\theta _{i}\theta _{j}=-\theta _{j}\theta _{i}$, hence $\theta
_{i}^{2}=0$. We denote such superalgebra with $k[t_{1},\dots ,t_{m},\theta
_{1},\dots ,\theta _{n}]$. The reader may safely think of such superalgebra
when we make our statements regarding commutative superalgebras.

If $A$ is the polynomial superalgebra, we have:
\begin{equation}
A_{0}=\left\{ f_{0}+\sum_{r\text{ even}}f_{I}\theta
_{I}\,|\,I=\{i_{1}<\ldots <i_{r}\}\right\} ,\qquad A_{1}=\left\{ \sum_{s%
\text{ odd}}f_{J}\theta _{J}\,|\,J=\{j_{1}<\ldots <j_{s}\}\right\} .
\end{equation}%
where we are using the multi-index notation and $f_{I},f_{J}\in
k[t_{1},\dots ,t_{n}]$ the ordinary polynomial algebra in the commuting
variables $t_{1},\dots ,t_{n}$.

Let $V$ be a vector space and $A$ a commutative superalgebra. We define:
\begin{equation}
V(A)=(A_{0}\otimes V_{0})\oplus (A_{1}\otimes V_{1}).
\end{equation}%
If $V=k^{p|q}$, we most immediately have
\begin{equation}
V(A)=\{(a_{1},\dots ,a_{p},\al_{1},\dots \al_{q})\,|\,a_{i}\in A_{0},\al%
_{j}\in A_{1}\}
\end{equation}

We define the \textit{$A$-points} of the
\textit{general linear supergroup} $\rGL%
(p|q)(A)$, as the parity preserving linear maps from $V(A)$ to itself. An
easy calculation shows that:
\begin{equation}\label{glfun}
\rGL(p|q)(A)=\left\{
\begin{pmatrix}
(a_{ij}) & (\al_{il}) \\
(\al_{kj}) & (a_{kl})%
\end{pmatrix}%
\,\Big|\,a_{ij},a_{kl}\in A_{0},\quad \alpha _{il},\alpha _{kj}\in
A_{1}\right\} 
\end{equation}%
where $1\leq i,j\leq p$, $p+1\leq k,l\leq p+q$ and $\det (a_{ij})$, $\det
(a_{kl})$ are invertible. This is an ordinary group with the matrix
multiplication. The super nature of this geometric object lies into the
anticommuting entries of its odd part, namely the $\al_{rs}$'s.

\medskip We can identify $\rGL(p|q)(A)$ with the group of superalgebra
morphisms from the superalgebra
\begin{equation}
k[\mathrm{GL}(p|q)]:=k[x_{ij},\xi _{kl}][\det (x_{ij})_{1\leq i,j\leq
p}^{-1},\det (x_{ij})_{p+1\leq i,j\leq p+q}^{-1}]
\end{equation}%
to the superalgebra $A$. Let us see this identification through an example
(the general case is a straightforward modification of it). Consider
\begin{equation}
\rGL(1|1)(A)=\left\{
\begin{pmatrix}
a & \al \\
\bet & b%
\end{pmatrix}%
\,\Big|\,a,b\in A_{0},\quad \al,\bet\in A_{1},\quad a,b\,\hbox{invertible}%
\right\}   \label{matrixgl}
\end{equation}%
and $k[\rGL(1|1)]=k[x,y,\xi ,\eta ][x^{-1},y^{-1}]$. A morphism $\phi
:k[x,y,\xi ,\eta ][x^{-1},y^{-1}]\lra A$ is determined by the images of the
generators, namely $\phi (x)=a$, $\phi (y)=b$, $\phi (\xi )=\al$, $\phi
(\eta )=\bet$, where $a$, $b$ are invertible in $A_{0}$ and $\al,\bet\in
A_{1}$. The identification of $\phi $ with a matrix in the form (\ref%
{matrixgl}) is then immediate.

\medskip The identification between $\rGL(p|q)(A)$ and the set of morphisms of
superalgebras as above, denoted by $\Hom_\salg(k[\rGL(p|q)],A)$, allows us
to say that the general linear supergroup is \textit{represented} by the
superalgebra $k[\rGL(p|q)]$. The information contained in $\rGL(p|q)(A)$
\textsl{for all} $A$ is effectively contained in the superalgebra $k[\rGL%
(p|q)]$. More appropriately, we call \textit{general linear supergroup over $%
k$} and we denote it by $\rGL(p|q)$, the functor that associates to a given
commutative superalgebra $A$ the group $\rGL(p|q)(A)$. The reader does not
need to be familiar with the theory of categories, but should be aware that
a supergroup functor $G$
is a way of giving, for any commutative superalgebra $A$, a group, denoted
by $G(A)$, that behaves nicely when we change $A$
(namely, if we have a morphism $A\lra B$,
this morphism should naturally induce another morphism $G(A) \lra G(B)$).
Furthermore, to fully deserve the name of supergroup,
the functor $G$ must be {\sl representable}, that is, there is a superalgebra
$k[G]$, playing the role of $k[\rGL(p|q)]$, so that we
can identify $G(A)$, the $A$-points of the
supergroup functor, with the morphisms $k[G] \lra A$.
However, for the
present work, we shall not be interested in these subtleties:
all of the supergroup
functors we consider in this paper are indeed representable.

\medskip
The \textit{reduced group} associated to a supergroup is
the ordinary group that we obtain by taking $A=k$. For example,
for $\rGL(p|q)$:
\begin{equation} \label{red-grp}
\rGL(p|q)(k)=\left\{
\begin{pmatrix}
(a_{ij}) & 0 \\
0 & (a_{kl})%
\end{pmatrix} \right\} \,= \, \rGL(p)\times \rGL(q)
\end{equation}
because the only value in a field $k$ that the nilpotent
variables $\al_{rs}$ can take is zero.


\medskip  At this point we need to make a step forward in this theory and
look at the differences 
in the choice of $k$. So to mark the difference between the different $k$'s,
we speak of $k$-supergroups or we say that a supergroup is defined over $k$.
For the purpose of the present paper, we need to consider $\C_s$-supergroups,
that we want to view as supergroups over $\R$.
Let us look at an
example and consider the supergroup  $\rGL(1|1)$ over $\C_{s}$;
again, the general case is not conceptually different. The superalgebra
representing the supergroup is
$\C_{s}[z,w,\zeta ,\eta ][z^{-1},w^{-1}]$ (see (\ref{matrixgl})).
This superalgebra will give us the $A$-points of $\rGL(1|1)$, when
$A$ is a $\C_s$-superalgebra, while now we want to determine the
$A$-points of $\rGL(1|1)$ as a {\sl real} supergroup, that is when
$A$ is a {\sl real} superalgebra. 
We then define the $A$-points of the $\C_s$-supergroup $\rGL(1|1)$, viewed as
$\R$-supergroup, the $A \otimes \C_s$ points of $\rGL(1|1)$:
$$
\rGL(1|1)_\R(A)=\rGL(1|1)(A \otimes \C_s)=\Hom_\salg(\C_{s}[z,w,\zeta ,\eta
][z^{-1},w^{-1}],A\otimes \C_s)
$$
where the tensor
product is over $\R$.
In fact, a morphism $\psi :\C_{s}[z,w,\zeta ,\eta
][z^{-1},w^{-1}]\lra A\otimes \C_{s}$ is specified once we know $\psi (z)$, $%
\psi (w)$, $\psi (\zeta )$, $\psi (\eta )$. Let us look at $\psi
(z)=a\otimes 1+b\otimes j$. The image of $z$ is effectively recovered
by the pair $(a,b)$ with $a,b\in A_{0}$. So we see that a complex
indeterminate $z$ is associated with two real indeterminates.
The images of the $4$
$\C_s$-generators $z,w.\psi,\zeta$ give $8$ elements of the real
algebra $A$, as one expects (in analogy to what
we expect for ordinary vector spaces or algebras: the complex coordinates double
their number, when viewed as real). 
For the $\C_s$ general linear supergroup:
\begin{equation} \label{cs-real}
(\rGL(p|q)(A))_{\mathbb{R}}=\Hom_{\salg}(\C_{s}[[x_{ij},\xi _{kl}][\det
(x_{ij})_{1\leq i,j\leq p}^{-1},\det (x_{ij})_{p+1\leq i,j\leq
p+q}^{-1}],A\otimes \C_{s})
\end{equation}
The definition for a generic $\C_s$-supergroup is: 
\begin{equation}
G_{\mathbb{R}}=\Hom_{\salg}(\C_{s}[G],A\otimes \C_{s})
\end{equation}


\begin{thebibliography}{999}
\bibitem{GL} Y. A. Gol'fand, E. P. Likhtman, \textit{Extension of the
Algebra of Poincar\'{e} Group and Violation of }$\mathit{P}$\textit{\
Invariance}, JETP Lett. \textbf{13}, 323 (1971).

\bibitem{VA} D. V. Volkov, V. P. Akulov, \textit{Is the Neutrino a Goldstone
Particle?}, Phys. Lett. \textbf{B46}, 109 (1973).

\bibitem{WZ} J. Wess, B. Zumino, \textit{Supergauge Transformations in Four
Dimensions}, Nucl. Phys. \textbf{B70}, 39 (1974).

\bibitem{SS-1} A. Salam, J. Strathdee, \textit{Super-Gauge Transformations},
Nucl. Phys. \textbf{B76}, 477 (1974).

\bibitem{SS-2} A. Salam, J. Strathdee, \textit{Unitary Representations of
Supergauge Symmetries}, Nucl. Phys. \textbf{B80}, 499 (1974).

\bibitem{FZ} S. Ferrara, B. Zumino, \textit{Supergauge Invariant Yang-Mills
Theories}, Nucl. Phys. \textbf{B79} (1974) 413.

\bibitem{FFVN} D. Z. Freedman, P. van Nieuwenhuizen, S. Ferrara, \textit{%
Progress Toward a Theory of Supergravity}, Phys. Rev. \textbf{D13} (1976)
3214-3218.

\bibitem{DZ} S. Deser, B. Zumino, \textit{Consistent Supergravity}, Phys.
Lett. \textbf{B62} (1976) 335.

\bibitem{GSW-book} M. B. Green, J. H. Schwarz, E. Witten : \textit{%
\textquotedblleft Superstring Theory"}, 2 Vols., Cambridge Monographs On
Mathematical Physics, Cambridge University Press (Cambridge), 1987.

\bibitem{Polchinski-book} J. Polchinski : \textit{\textquotedblleft String
theory"}, 2 Vols., Cambridge University Press (Cambridge), 1998.

\bibitem{vsv} V.~S.~Varadarajan : \textit{\textquotedblleft Supersymmetry
for Mathematicians: An Introduction"}, Courant Lecture Notes \textbf{1},
AMS, 2004.

\bibitem{bcf1} L. Balduzzi, C. Carmeli, R. Fioresi, \textit{The local
functor of points of supermanifolds}, Expositiones Mathematic\ae\ \textbf{28}
(2010), 201-217, \texttt{arXiv:0908.1872 [math.RA]}.

\bibitem{bcf2} L. Balduzzi, C. Carmeli, R. Fioresi, \textit{A Comparison of
the functors of points of Supermanifolds}, J. Algebra Appl. \textbf{12},
(2013), 1407-1415, \texttt{arXiv:0902.1824 [math.RA]}.

\bibitem{sh} A.~S. Shvarts, \textit{On the definition of superspace},
Teoret. Mat. Fiz. \textbf{60}(1):37-42 (1984).

\bibitem{vo} A. Voronov, \textit{Maps of supermanifolds}, Teoret. Mat. Fiz.
\textbf{60}(1):43-48 (1984).

\bibitem{be} F.~A.~Berezin : \textquotedblleft \textit{Introduction to
superanalysis}", D.~Reidel Publishing Company, Holland, 1987.

\bibitem{ma1} Y.~I.~Manin : \textquotedblleft \textit{Topics in non
commutative geometry"}, Princeton University Press, 1991.
-
\bibitem{ma2} Y.~I.~Manin : \textquotedblleft \textit{Gauge field theory and
complex geometry"}, translated by N. Koblitz, J.R. King, Springer-Verlag,
Berlin-New York, 1988.

\bibitem{FL-1} R. Fioresi, E. Latini, \textit{The symplectic origin of
conformal and Minkowski superspaces}, J.Math.Phys. \textbf{57} no.2, 022307 (2016), \texttt{arXiv:1506.09086 [hep-th]}.

\bibitem{ccf} C. ~Carmeli, L.~Caston, R.~Fioresi : \textquotedblleft \textit{%
Mathematical Foundation of Supersymmetry"}, with an appendix with I.
Dimitrov, EMS Ser. Lect. Math., European Math. Soc., Zurich, 2011.

\bibitem{flv} R. Fioresi, M. A. Lled\'{o}, V. S. Varadarajan, \textit{The
Minkowski and conformal superspaces}, J. Math. Phys. \textbf{48}, 113505,
(2007), \texttt{math/0609813 [math.RA]}.

\bibitem{Hurwitz} A. Hurwitz, \textit{\"{U}ber die Composition der
quadratischen Formen von beliebig vielen Variabeln}, Nachr. Ges. Wiss. G{%
\"{o}}ttingen (1898), 309-316.

\bibitem{Baez-Huerta-1} J. C. Baez, J. Huerta, \textit{Division Algebras and
Supersymmetry I}, in : \textit{\textquotedblleft Superstrings, Geometry,
Topology, and }$\mathit{C}^{\ast }$\textit{-algebras"}, eds. R. Doran, G.
Friedman and J. Rosenberg, Proc. Symp. Pure Math. \textbf{81}, AMS,
Providence, 2010, 65-80, \texttt{arXiv:0909.0551 [hep-th]}.

\bibitem{cederwall1} M.~Cederwall, \textit{Jordan algebra dynamics},
Nucl.Phys. \textbf{B302} (1988) 81.

\bibitem{cederwall2} M.~Cederwall, \textit{Octonionic particles and the }$%
S(7)$\textit{\ symmetry}, J.Math.Phys. \textbf{33} (1992) 388.

\bibitem{cederwall3} M.~Cederwall, \textit{Introduction to Division
Algebras, Sphere Algebras and Twistors,} \texttt{arXiv:hep-th/9310115}.

\bibitem{evans} J. M. Evans, \textit{Supersymmetric Yang-Mills theories and
division algebras,} Nucl. Phys. \textbf{B298} (1988), 92-108.

\bibitem{Huerta-2} J. C. Baez, J. Huerta, \textit{Division Algebras and
Supersymmetry II}, Adv. Theor. Math. Phys. \textbf{15} (2011) 5, 1373-1410,
\texttt{arXiv:1003.3436 [hep-th]}.

\bibitem{Huerta-3} J. Huerta, \textit{Division Algebras and Supersymmetry III%
}, Adv. Theor. Math. Phys. \textbf{16} (2012) 5, 1485-1589, \texttt{%
arXiv:1109.3574 [hep-th]}.

\bibitem{Huerta} J. Huerta, \textit{Division Algebras and Supersymmetry IV},
\texttt{arXiv:1409.4361 [hep-th]}.

\bibitem{Bern} Z.~Bern, J.~J.~M.~Carrasco, H.~Johansson, \textit{%
Perturbative Quantum Gravity as a Double Copy of Gauge Theory}, Phys.\ Rev.\
Lett.\ \textbf{105}, 061602 (2010), \texttt{arXiv:1004.0476 [hep-th]}.

\bibitem{ICL-1309} A. Anastasiou, L. Borsten, M.J. Duff, L.J. Hughes, S.
Nagy, \textit{Super Yang-Mills, division algebras and triality}, JHEP
\textbf{1408} (2014) 080, \texttt{arXiv:1309.0546 [hep-th]}.

\bibitem{ICL-2} A. Anastasiou, L. Borsten, M.J. Duff, L.J. Hughes, S. Nagy,
\textit{Yang-Mills origin of gravitational symmetries}, Phys. Rev. Lett.
\textbf{113} (2014) no.23, 231606, \texttt{arXiv:1408.4434 [hep-th]}.

\bibitem{ICL-3} A. Anastasiou, L. Borsten, M.J. Hughes, S. Nagy, \textit{%
Global symmetries of Yang-Mills squared in various dimensions}, JHEP \textbf{%
1601} (2016) 148, \texttt{arXiv:1502.05359 [hep-th]}.

\bibitem{MS} H. Freudenthal, \textit{Lie groups in the foundations of
geometry}, Adv. Math. \textbf{1}, 145-190 (1964).

\bibitem{MS2} J. Tits, \textit{Alg\'{e}bres Alternatives, Alg\'{e}bres de
Jordan et Alg\'{e}bres de Lie Exceptionnelles}, Indag. Math. \textbf{28},
223-237 (1966).

\bibitem{BS-2} C.H. Barton, A. Sudbery, \textit{Magic squares and matrix
models of Lie algebras, }Adv. in Math. \textbf{180} (2003), 596-647, \texttt{%
arXiv:math/0203010 [math.RA]}.

\bibitem{Baez} J. C. Baez, \textit{The Octonions}, Bull. Amer. Math. Soc.
\textbf{39}, 145-205 (2002), \texttt{arXiv:math/0105155 [math.RA]}.

\bibitem{vsv-2} V.~S.~Varadarajan : \textquotedblleft \textit{Lie groups,
Lie algebras, and their representations}", Graduate Text in Mathematics.
Springer-Verlag, New York, 1984.

\bibitem{BS-1} C.H. Barton, A. Sudbery, \textit{Magic Squares of Lie Algebras%
}, \texttt{arXiv:math/0001083 [math.RA]}.

\bibitem{MS-2-Groups} T. Dray, J. Huerta, J. Kincaid, \textit{The Magic
Square of Lie Groups: The }$2\times 2$\textit{\ Case}, Lett. Math. Phys.
\textbf{104} (2014) 1445-1468.

\bibitem{Dray-2} T. Dray, C.A. Manogue, R.A. Wilson, \textit{A symplectic
representation of }$E_{7}$, Comment. Math. Univ. Carolin. \textbf{55},
387-399 (2014), \texttt{arXiv:1311.0341 [math.RA]}.

\bibitem{Dray-3} J.~Kincaid, T.~Dray, \textit{Division algebra
representations of }$\mathit{SO(4,2)}$, Mod.\ Phys.\ Lett.\ \textbf{A29},
no. 25, 1450128 (2014), \texttt{arXiv:1312.7391 [math.RA]}.

\bibitem{Dray-4} T.~Dray, C.~A.~Manogue, \textit{Octonionic Cayley Spinors
and }$\mathit{E}_{6}$, Comment. Math. Univ. Carolin. \textbf{51}, 193-207
(2010), \texttt{arXiv:0911.2255 [math.RA]}.

\bibitem{tractor} T.N. Bailey, M.G. Eastwood, A.R. Gover, \textit{Thomas's
structure bundle for conformal, projective and related structures}, Rocky
Mountain J. Math. \textbf{24} (1994), 1191-1217.

\bibitem{Gover} A.~R.~Gover, A.~Shaukat, A.~Waldron, \textit{Tractors, Mass
and Weyl Invariance}, Nucl.\ Phys.\ \textbf{B812}, 424 (2009), \texttt{%
arXiv:0810.2867 [hep-th]}.

\bibitem{Curry} S.~Curry, A.~R.~Gover, \textit{An introduction to conformal
geometry and tractor calculus, with a view to applications in general
relativity}, \texttt{arXiv:1412.7559 [math.DG]}.

\bibitem{RodGover:2012ib} A.~Rod Gover, E.~Latini, A.~Waldron, \textit{Poincar\'{e}-Einstein Holography for Forms via Conformal Geometry in the Bulk}, Memoirs of the American Mathematical Society \textbf{235} (2015), \texttt{aarXiv:1205.3489 [math.DG]}.

\bibitem{tractorparabolic} A. \v{C}ap, A.R. Gover, \textit{Tractor bundles
for irreducible parabolic geometries}, Global analysis and harmonic
analysis, S{\'{e}}min. Congr. \textbf{4}, 129, Soc. Math. France 2000.

\bibitem{FG} C.~ Fefferman, C.R.~ Graham, \textit{Conformal invariants}, in
: \textit{\textquotedblleft The mathematical heritage of Cartan"} (Lyon,
1984). Asterisque 1985, Numero Hors Serie, 95-116.

\bibitem{Kuzenko} S.~M.~Kuzenko, \textit{Conformally compactified Minkowski
superspaces revisited}, JHEP \textbf{1210}, 135 (2012), \texttt{%
arXiv:1206.3940 [hep-th]}.

\bibitem{Klemm-Nozawa} D. Klemm, M. Nozawa, \textit{Geometry of Killing
spinors in neutral signature}, Class. Quant. Grav. \textbf{32} (2015) no.18,
185012, \texttt{arXiv:1504.02710 [hep-th]}.

\bibitem{Hull-1} C. M. Hull, \textit{Duality and the signature of space-time}%
, JHEP \textbf{9811} (1998) 017, \texttt{hep-th/9807127}.

\bibitem{Hull-2} C. M. Hull, \textit{Timelike T-Duality, de Sitter Space,
Large }$\mathit{N}$\textit{\ Gauge Theories and Topological Field Theory},
JHEP \textbf{9807} (1998) 021, \texttt{hep-th/9806146}.

\bibitem{Ferrara-Spinors} S. Ferrara, \textit{Spinors, superalgebras and the
signature of space-time}, \texttt{hep-th/0101123}.

\bibitem{Bryant-1} R. L. Bryant, \textit{Pseudo-Riemannian metrics with
parallel spinor fields and vanishing Ricci tensor}, in : \textit{%
\textquotedblleft Global analysis and harmonic analysis"} (Marseille-Luminy,
1999), vol. \textbf{4} of S\'{e}min. Congr., pp. 53--94. Soc. Math. France,
Paris, 2000, \texttt{math/0004073 [math.DG]}.

\bibitem{Dun-1} M. Dunajski, \textit{Anti-self-dual four manifolds with a
parallel real spinor}, Proc. Roy. Soc. Lond. \textbf{A458} (2002) 1205,
\texttt{math/0102225 [math.DG]}.

\bibitem{Dun-2} M. Dunajski, \textit{Einstein-Maxwell-dilaton metrics from
three-dimensional Einstein-Weyl structures}, Class. Quant. Grav. \textbf{23}
(2006) 2833, \texttt{gr-qc/0601014}.

\bibitem{Dun-3} M. Dunajski, S. West, \textit{Anti-self-dual conformal
structures in neutral signature}, \texttt{math/0610280 [math.DG]}.

\bibitem{Hervik} S. Hervik, \textit{Pseudo-Riemannian VSI spaces II}, Class.
Quant. Grav. \textbf{29}, 095011 (2012), \texttt{arXiv:1504.01616 [math-ph]}.

\bibitem{OV} H. Ooguri, C. Vafa, \textit{Selfduality and }$\mathcal{N}%
\mathit{=2}$\textit{\ string magic}, Mod. Phys. Lett. \textbf{A5} (1990)
1389.

\bibitem{Penrose} R. Penrose, \textit{Twistor algebra}, J. Math. Phys.
\textbf{8} (1967) 345.

\bibitem{Witten} E. Witten, \textit{Perturbative gauge theory as a string
theory in twistor space}, Commun. Math. Phys. \textbf{252} (2004) 189,
\texttt{hep-th/0312171}.

\bibitem{Rios} M. Rios, \textit{Extremal Black Holes as Qudits}, \texttt{%
arXiv:1102.1193 [hep-th]}.

\bibitem{Nahm} W.~Nahm, \textit{Supersymmetries and their representations},
Nucl. Phys. \textbf{B135} (1978) 149.

\bibitem{flmink} R. Fioresi, M. A. Lled\'{o} : \textit{\textquotedblleft The
Minkowski and Conformal Superspaces: The Classical and Quantum Descriptions"}%
, World Scientific Publishing, 2015.

\bibitem{Sudbery} A. Sudbery, \textit{Division Algebras, (Pseudo)Orthogonal
Groups and Spinors}, J. Phys. \textbf{A17}, 939 (1984).

\bibitem{Kugo-Townsend} T. Kugo, P. K. Townsend, \textit{Supersymmetry and
the Division Algebras}, Nucl. Phys. \textbf{B221} (1983) 357.

\bibitem{cfln} D. Cervantes, R. Fioresi, M.A. Lled\'{o}, F. Nadal, \textit{%
Quadratic deformation of Minkowski space}, Fortschr. Phys. \textbf{60}
(2012), no. 9-10, 970-976. 53Cxx (32L25).

\bibitem{cfl} D. Cervantes, R. Fioresi, M.A. Lled\'{o},\emph{\ }\textit{The
quantum chiral Minkowski and conformal superspaces}, Adv. Theor. Math. Phys.
\textbf{15} (2011), no. 2, 565-620, \texttt{arXiv:1007.4469 [math.QA]}.

\bibitem{cfl2} D. Cervantes, R. Fioresi, M.A. Lled\'{o} : \textit{%
\textquotedblleft On chiral quantum superspaces"}, in :\textit{\
\textquotedblleft Supersymmetry in Mathematics and Physics"}, 69-99, Lecture
Notes in Math. \textbf{2027}, Springer, Heidelberg, 2011.

\bibitem{fquant} R. Fioresi, \textit{Quantizations of flag manifolds and
conformal space time}, Rev. Math. Phys. \textbf{9} (1997), no. 4, 453-465.

\bibitem{fquant2} R. Fioresi, \textit{Quantum deformation of the flag variety%
}, Comm. Algebra \textbf{27} (1999), no. 11, 5669-5685.

\bibitem{hypercomplex} I.L. Kantor, A.S. Solodovnikov : \textit{%
\textquotedblleft Hypercomplex Numbers: An Elementary Introduction to
Algebras"}, Springer, New York, 1983.

\bibitem{split-H} K. Carmody, \textit{Circular and hyperbolic quaternions,
octonions, sedionions}, Applied Mathematics and Computation \textbf{84}(1)
(1997), 27--47.

\bibitem{Gun-2} M. G\"{u}naydin, O. Pavlyk, \textit{Spectrum Generating
Conformal and Quasiconformal }$\mathit{U}$\textit{-Duality Groups,
Supergravity and Spherical Vectors}, JHEP \textbf{1004} (2010) 070, \texttt{%
arXiv:0901.1646 [hep-th]}.

\bibitem{Tray-Manogue-Book} T. Dray, C. A. Manogue : \textit{%
\textquotedblleft The Geometry of the Octonions"}, World Scientific, 2015.

\bibitem{Springer-book} T. Springer, F. D. Veldkamp : \textit{%
\textquotedblleft Octonions, Jordan Algebras and Exceptional Groups"},
Springer, 2013.

\bibitem{Evans} J. M. Evans, \textit{Trialities and Exceptional Lie
Algebras: Deconstructing the Magic Square}, \texttt{arXiv:0910.1828 [hep-th]}%
.

\bibitem{CFMZ1-D=5} B.L. Cerchiai, S. Ferrara, A. Marrani, B. Zumino,
\textit{Charge Orbits of Extremal Black Holes in Five Dimensional
Supergravity}, Phys. Rev. \textbf{D82} (2010) 085010, \texttt{%
arXiv:1006.3101 [hep-th]}.

\bibitem{Magic-Coset-Decomp} S.L. Cacciatori, B.L. Cerchiai, A. Marrani,
\textit{Magic Coset Decompositions}, Adv. Theor. Math. Phys. \textbf{17}
(2013) 1077-1128, \texttt{arXiv:1201.6314 [hep-th]}.

\bibitem{ADFMT-1} L. Andrianopoli, R. D'Auria, S. Ferrara, A. Marrani, M.
Trigiante, \textit{Two-Centered Magical Charge Orbits}, JHEP \textbf{1104}
(2011) 041, \texttt{arXiv:1101.3496 [hep-th]}.

\bibitem{McCrimmon} K. McCrimmon : \textit{\textquotedblleft A Taste of
Jordan Algebras"}, Springer, 2004.

\bibitem{Iordanescu} R. Iordanescu : \textit{\textquotedblleft Jordan
structures in Analysis, Geometry and Physics"}, Editura Academiei Rom\^{a}%
ne, 2009.

\bibitem{Jacobson} N. Jacobson, \textit{Structure and representations of
Jordan algebras}, American Mathematical Society Colloquium Publications,
Vol. XXXIX. American Mathematical Society, Providence, R.I., 1968.

\bibitem{JWVN} P.~Jordan, J.~von Neumann, E.P. Wigner, \textit{On an
algebraic generalization of the quantum mechanical formalism}, Ann. Math.
\textbf{35} (1934) no. 1, 29--64.

\bibitem{Budinich-1} P. Budinich, \textit{From the Geometry of Pure Spinors
with their Division Algebras to Fermion's Physics}, Found. Phys. \textbf{32}
(2002) 1347-1398, \texttt{arXiv:hep-th/0107158}.

\bibitem{Budinich-2} P. Budinich, \textit{Internal Symmetry From Division
Algebras in Pure Spinor Geometry}, Proceedings of Institute of Mathematics
of NAS of Ukraine 2004, Vol. \textbf{50}, Part 2, 654--665, \texttt{%
arXiv:hep-th/0311045}.

\bibitem{Charlton-Th} P. Charlton : \textit{\textquotedblleft The geometry
of pure spinors, with applications"}, PhD thesis, University of Newcastle,
Dept. of Mathematics, 1997.

\bibitem{Spinor-Algebras} R. D'Auria, S. Ferrara, M.A. Lled\'{o}, V.S.
Varadarajan, \textit{Spinor algebras}, J. Geom. Phys. \textbf{40} (2001)
101-128, \texttt{hep-th/0010124}.

\bibitem{Cartan} \'{E}. Cartan : \textit{\textquotedblleft Le\c{c}ons sur la
Theorie des Spineurs"}, Paris, Hermann, 1937.

\bibitem{Chevalley} C. Chevalley : \textit{\textquotedblleft The algebraic
theory of spinors"}, New York, Columbia University Press, 1954.

\bibitem{Berkovits} N. Berkovits, \textit{Super-Poincar\'{e} Covariant
Quantization of the Superstring}, JHEP \textbf{0004}, 018 (2000), \texttt{%
arXiv:hep-th/0001035}.

\bibitem{PSF} N. Berkovits, \textit{ICTP Lectures on Covariant Quantization
of the Superstring}, \textit{arXiv:hep-th/0209059}.

\bibitem{Igusa} J. Igusa, \textit{A classification of spinors up to
dimension twelve}, Am. J. of Math. \textbf{92} (1970), 997--1028.

\bibitem{KV78} V.G. Kac, E.B. Vinberg, \textit{Spinors of 13-dimensional
space}, Adv. in Math. \textbf{30} (1978), 137--155.

\bibitem{Pop80} V.L. Popov, \textit{Classification of spinors of dimension
fourteen}, Trans. Moscow Math. Soc. \textbf{1} (1980), 181--232.

\bibitem{Zhu92} X.-W. Zhu, \textit{The classification of spinors under }$%
\mathit{GSpin}_{14}$\textit{\ over finite fields}, Trans. Am. Math. Soc.
\textbf{333} (1992), no. 1, 95--114.

\bibitem{AE82} L.V. Antonyan, A.G. \`{E}lashvili, \textit{Classification of
spinors in dimension sixteen}, Trudy Tbiliss. Mat. Inst. Razmadze Akad. Nauk
Gruzin. SSR \textbf{70} (1982), 4--23.

\bibitem{Pure-Spinors-Polish} S. Giler, P. Kosinski, J. Rembielinski,
\textit{On }$\mathit{SO(p,q)}$\textit{\ Pure Spinors}, Acta Phys. Pol. Vol.
\textbf{B18}, no. 8, 713 (1987).

\bibitem{Furlan} P. Furlan, R. Raczka, \textit{Nonlinear spinor
representations}, J. Math. Phys. \textbf{26}, 3021-3032 (1985).

\bibitem{Bryant-2} R. L. Bryant, \textit{Remarks on spinors in low dimension}%
, Unpublished notes, 1999.

\bibitem{GST} M. G\"{u}naydin, G. Sierra and P. K. Townsend, \textit{%
Exceptional Supergravity Theories and the Magic Square}, Phys. Lett. \textbf{%
B133}, 72 (1983). M. G\"{u}naydin, G. Sierra and P. K. Townsend, \textit{The
Geometry of }$\mathcal{N}\mathit{=2}$\textit{\ Maxwell-Einstein Supergravity
and Jordan Algebras}, Nucl. Phys. \textbf{B242}, 244 (1984).

\bibitem{Gunaydin-D=6} M. G\"{u}naydin, H. Samtleben, E. Sezgin, \textit{On
the Magical Supergravities in Six Dimensions}, Nucl. Phys. \textbf{B848}
(2011) 62-89, \texttt{arXiv:1012.1818 [hep-th]}.

\bibitem{cfk1} C.~Carmeli, R.~Fioresi, S. D. Kwok, \textit{SUSY structures,
representations and Peter-Weyl theorem for $S^{1|1}$}, J. Geom. Phys.
\textbf{95}, 144-158 (2015), \texttt{arXiv:1407.2706 [math.RT]}.

\bibitem{cfk2} C.~Carmeli, R.~Fioresi, S. D. Kwok, The \textit{Peter-Weyl
theorem for ${SU}(1|1)$}, P-Adic Numbers, Ultrametric Analysis, and
Applications, Vol. \textbf{7}, no. 4, 266-275 (2015), \texttt{%
arXiv:1509.07656 [math.RT]}.
\end{thebibliography}
\end{document}